\documentclass[preprintnumbers,amsmath,amssymb,superscriptaddress]{revtex4}

\usepackage{bm}
\usepackage{graphicx}

\begin{document}

\title{Chiral symmetry breaking via crystallization
of the glycine and $\alpha$-amino acid system: a mathematical model}
\author{Celia Blanco}
\email{blancodtc@cab.inta-csic.es} \affiliation{Centro de
Astrobiolog\'{\i}a (CSIC-INTA), Carretera Ajalvir Kil\'{o}metro 4,
28850 Torrej\'{o}n de Ardoz, Madrid, Spain}
\author{David Hochberg}
\email{hochbergd@cab.inta-csic.es} \affiliation{Centro de
Astrobiolog\'{\i}a (CSIC-INTA), Carretera Ajalvir Kil\'{o}metro 4,
28850 Torrej\'{o}n de Ardoz, Madrid, Spain}

\begin{abstract}
 We introduce and numerically solve a
mathematical model of the experimentally established mechanisms
responsible for the symmetry breaking transition observed in the
chiral crystallization experiments reported by I. Weissbuch, L.
Addadi, L. Leiserowitz and M. Lahav, J. Am. Chem. Soc. \textbf{110}
(1988), 561-567. The mathematical model is based on five basic
processes: (1) The formation of achiral glycine clusters in
solution, (2) The nucleation of oriented glycine crystals at the
air/water interface in the presence of hydrophobic amino acids, (3)
A kinetic orienting effect which inhibits crystal growth, (4) The
enantioselective occlusion of the amino acids from solution, and (5)
The growth of oriented host glycine crystals at the interface. We
translate these processes into differential rate equations. We first
study the model with the orienting process (2) without (3) and then
combine both allowing us to make detailed comparisons of both
orienting effects which actually act in unison in the experiment.
Numerical results indicate that the model can yield a high
percentage orientation of the mixed crystals at the interface and
the consequent resolution of the initially racemic mixture of amino
acids in solution. The model thus leads to separation of
enantiomeric territories, the generation and amplification of
optical activity by enantioselective occlusion of chiral additives
through chiral surfaces of glycine crystals.
\end{abstract}

\maketitle

\section{Introduction}

Theoretical proposals for prebiotic chemistry suggest that
homochirality emerged in nature in abiotic times via deterministic
or chance mechanisms \cite{Guijarro2009}. The abiotic scenario for
the emergence of single homochirality in the biological world
implies that single asymmetry could have emerged provided a small
fluctuation from the racemic state can be amplified to a state
useful for biotic evolution. For this reason, experiments that can
demonstrate the feasibility of stochastic mirror symmetry breaking
involving the self assembly of molecular clusters, and in possible
conjunction with interface effects, are particularly important. This
is because, once generated by chance or an initial chiral
fluctuation, the chirality can then be preserved and transmitted to
the rest of the system provided that the symmetry breaking step is
coupled to a sequential step of efficient amplification via
self-replication reactions. Some relevant features common to such
systems are that they take into account the small fluctuations about
the racemic state and that they display nonlinear kinetic effects.
Stochastic scenarios are theoretically well
understood\cite{Konde1985,Avetisov1987} and are experimentally
feasible in the laboratory \cite{Lahav2005}.

Among the experiments dedicated to exploring chance mechanisms in
chirality, a particularly noteworthy and important result with a
marked relevance for prebiotic chemistry stands out. The experiment
we are interested in modeling here was reported some years ago by
the Rehovot group which dealt with an autocatalytic process for the
resolution of racemic $\alpha$-amino acids within crystals of
glycine grown at the air/solution interface
\cite{Lahav1984,Lahav1988}. They applied cooperative crystallization
processes for the spontaneous separation of racemic mixtures of
$\alpha$-amino acids rich with glycine into optically pure
enantiomers. Their experimental model involves slow evaporation of
aqueous solutions of the centrosymmetric form of glycine containing
racemic mixtures of $\alpha$-amino acids. Due to the unique crystal
structure of glycine, all chiral D-amino acids except for proline
are occluded within the crystal through the $(010)$ face, whereas
the L-amino acids are occluded through the $(0\bar 1 0)$ face.
Glycine crystals float in solution, so that only one face is
available for growth. Thus when glycine crystals are grown at the
air/water interface in the presence of DL-amino acids, only one of
its enantiotopic faces, say $(010)$, is exposed to the solution and
so picks up only the D-amino acid together with glycine. By
symmetry, crystals exposing their $(0{\bar 1}0)$ face towards
solution occlude only the L-enantiomers. Now, if by \textit{chance}
a single or small number of oriented glycine crystals grow initially
at the interface, the bulk solution will be enhanced with the amino
acid of one handedness. The preservation and transmission of the
chirality generated by chance of the original
\textit{Adam}\footnote[5]{In the original Hebrew version of the
Bible, \textit{adam} stands for all members of mankind: man, woman,
boy or girl. So there are two "Adams" in the Hebrew language, (i)
Adam denoting Mankind and so implicitly including \textit{both
sexes}, as distinct from (ii) Adam the male, who was differentiated
from Eve the female. We use the first definition as a metaphor for
the formation of the first glycine crystal, where both potential
interface orientations are implicit, instead of Eve, which might
represent a secondary crystallization. } crystal means that new
crystals grown at later stages at this interface must adopt the same
orientation. There are two proven ways this is
achieved\cite{Lahav1984,Lahav1988}: by means of an (i) hydrophobic
effect and (ii) by a kinetic inhibition effect. Regarding the first
effect, if the solution contains hydrophobic amino acids, these tend
to accumulate at the interface forming two-dimensional domains
acting as templates for the oriented crystallization of the glycine
crystals. Thus, the L-amino acids induce crystallization of floating
glycine crystals exposing their $(010)$ face towards solution and
these occlude only the D-amino acids. This asymmetric induction has
been established experimentally\cite{Lahav1984,Lahav1989}. As for
the second effect, this comprises an enantioselective inhibition of
the glycine nuclei by the amino acids present in solution (these can
be both partially dissolved hydrophobic as well as hydrophilic amino
acids). This independent effect was proven experimentally by
achieving complete orientation of the floating glycine nuclei when
grown in the presence of DL-leucine and hydrophilic L amino acids.
The presence of the DL-leucine is to ensure nucleation of floating
glycine crystals exposing either enantiotopic face towards solution
in a $1:1$ ratio whereas increasing the concentration of the
hydrophilic amino acid additives inhibited the glycine nuclei
exposing their $(0{\bar 1}0)$ face and so prevented their further
growth. These hydrophilic amino acids inhibit the crystal nuclei
from growing and developing into macroscopic
crystals\cite{Lahav2003}. Both the hydrophobic and kinetic effects
act cooperatively in the same direction in that they both contribute
to the territorial segregation of the enantiomers.

\begin{figure}[h]
\centering
  \includegraphics[width=0.45\textwidth]{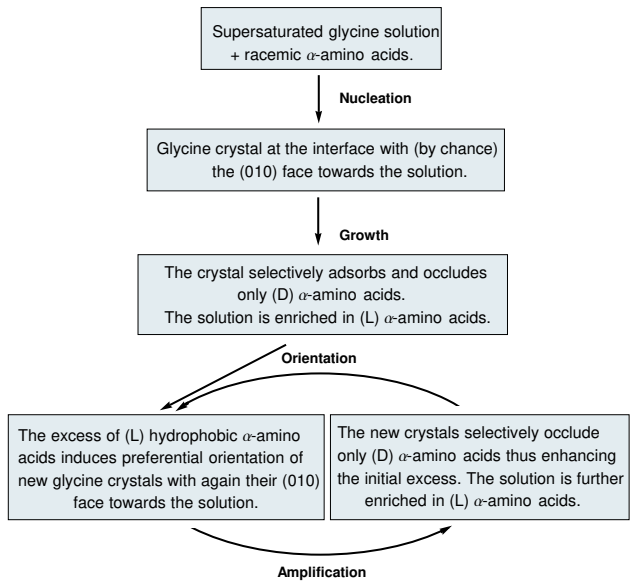}
  \caption{The overall process of mirror symmetry breaking and amplification.
  Scheme adapted from references\cite{Lahav1988,Lahav2005}.}
  \label{SchemeII}
\end{figure}

The overall experimental process can be summarized by the following
steps (Figure \ref{SchemeII})\cite{Lahav1988},\cite{Lahav2005}:(i)
mirror symmetry breaking via the enantioselective occlusion of one
of the enantiomers of the racemic $\alpha$-amino acids within
crystals of glycine grown at the air/solution interface, (ii) the
oriented crystal formed at the interface operates as a seed for
further occlusion of amino acids of the same handedness, (iii) the
amplification experiments comprises self-aggregation of hydrophobic
$\alpha$-amino acids into chiral clusters that operate as templates
for an oriented crystallization of fresh crystals of glycine. (i.e,
the hydrophobic effect), and (iv) enantioselective inhibition of
embryonic nuclei of glycine generated at the air/solution interface
by the water soluble $\alpha$-amino acids formed in excess during
the process. (i.e., the kinetic effect).
The fundamental relevance of this experiment is that it provides a
simple chemical model for the generation and amplification of
optically active amino acids in prebiotic conditions. They used a
centrosymmetric crystal structure of glycine, which is the only
achiral amino acid and is the major component found in modern
prebiotic synthesis experiments\cite{Ruiz2010}. The possibility to
maintain and propagate the same chirality to the entire system can
lead in principle to a complete separation of enantiomeric
territories and subsequent optical activity. The enantiomeric
resolution and symmetry breaking is achieved without the need to
input mechanical energy: in particular there is no need for stirring
\cite{Konde1990} nor grinding \cite{Viedma2005}, and it involves
only amino acids which have an immediate and obvious relevance for
prebiotic chemistry. That the experiment dispenses with the need for
mechanical energy is a most remarkable feature. Some provisional
explanations of how this is achieved are offered in the final
conclusions.

In spite of its importance, we are unaware of any prior attempts to
offer a mathematical model for describing the reported phenomenon.
The purpose of this paper is to provide a simple and minimal
mathematical model of the key processes to further elucidate the
mechanisms responsible for the observed symmetry breaking and to
gain further understanding of the experiment itself. Our aim is to
describe the resolution process by means of a kinetic mean field
model in which we include only the minimal essentials responsible
for the crystal growth and the symmetry breaking phenomenon. The
mathematical model is based on five basic processes that parallel
closely those that have been established experimentally: (1) The
formation of achiral glycine clusters in solution, (2) The
nucleation of oriented glycine crystals at the air/water interface
in the presence of hydrophobic amino acids, (3) A kinetic orienting
effect which inhibits crystal growth from embryonic crystals, (4)
the enantioselective occlusion of the amino acids from solution, and
(5) the growth of oriented host glycine crystals at the interface
via glycine up-take. We first introduce the model with only the
hydrophobic orienting process (2) as an approximation to the
complete model. This is a justifiable first approximation to
consider given that the experimental results prove that this effect
plays a \textit{dominant} role in the orientation of the glycine
crystals \cite{Lahav1988}. The major simplifications are introduced
at this stage and we use this model to prove that the symmetric
state is unstable. We then include both effects (2) and (3) in a
complete final version allowing us to make detailed comparisons of
both effects which actually act in unison in the experiment. We
present numerical studies of the effects of varying amounts of both
the hydrophobic and hydrophilic additives on the orientation of the
floating glycine crystals, as was originally considered in the
experiments. The results indicate that our simple model can yield a
high percentage orientation of the glycine-host plus
guest-enantiomer mixed crystals at the interface with the consequent
resolution of the initially racemic mixture of amino acids in
solution. The mathematical model thus leads to separation of
enantiomeric territories, the generation and amplification of
optical activity by enantioselective occlusion of chiral additives
through chiral surfaces of glycine crystals.

\section{Reaction model}

\subsection{Preliminaries}

A full experimental account is given in the two papers
\cite{Lahav1984,Lahav1988} and is reviewed\cite{Lahav2005,Lahav2011}
together the references therein; and these serve as basic
inspiration and background guide for elaborating the model we
present in this paper.

Following the experimental conditions, we consider a supersaturated
solution of glycine and racemic $\alpha$ amino acids. The glycine
monomers and clusters in solution will be denoted by $A_1$ and
$A_r$, respectively, the cluster being made up from $r$-monomer
units. It is reasonable to assume that a typical glycine cluster
size should exist $<r>=M$ for $M$ monomeric units. This is the
average size of the primary nucleation or seed crystal which can
then precipitate at the air/water interface where it will, in the
\textit{presence} of hydrophobic $D$ and $L$-amino acids, assume one
of two unique orientations \cite{Lahav1988}.  Let $X_M$ and $Y_M$
denote the two possible orientations of the glycine crystal floating
at the interface: $X$ stands for the $(010)$ enantiotopic face
exposed to solution and $Y$ for the other $(0{\bar 1}0)$
enantiotopic face exposed to solution. The $D$ and $L$ denote the
two amino acid enantiomers. For a strictly racemic solution of the
amino acids, (which is of course impossible to realize
experimentally\cite{Mislow2003}), we would expect an equal $X:Y =
1:1$ proportion of the two glycine crystal orientations at the
interface. However, if for example there is an imbalance in the
initial concentrations, say $L>D$, then more of the $Y$-oriented
crystals will be precipitated. Likewise for the $X$-oriented
crystals if instead $L<D$.

Next, we consider the  enantioselective occlusion of the amino acid
into the host glycine crystals: recall only the $(010)$ face of the
glycine crystal can occlude the $D$ amino acid whereas only the
$(0{\bar 1}0)$ face can occlude the $L$ amino acid; see Figure
\ref{SchemeII}. It will prove instructive to first write down a kind
of ``microscopic" model which contemplates all kinds of crystal
aggregations growing from both glycine take-up and the amino acid
occlusion and with cluster size dependent rates and then simplify by
taking the reaction rates independent of cluster size, and assuming
only cluster-monomer aggregation. We will not consider fragmentation
in any case: we assume irreversible steps from the outset. Since the
hydrophobic effect dominates over the kinetic one\cite{Lahav1988},
we will consider this one first.

\subsection{Model based on the hydrophobic crystal orientation effect}

We now introduce the explicit processes to be included in the
mathematical model. We transcribe the scheme in Figure
\ref{SchemeII} into reaction steps. Simplifications will follow
later on in order to obtain what we will consider as a minimal model
leading to mirror symmetry breaking.

\textsl{ The formation of glycine aggregates/clusters in solution}.
The achiral molecules of glycine in solution combine pairwise to
yield achiral glycine clusters in solution, where $r,s$ denote
number of monomers in the cluster and $\delta_{r,s}$ is the cluster
size dependent rate constant:
\begin{equation}\label{gcluster}
A_r  + A_s \stackrel{\delta_{r,s}}{\longrightarrow} A_{r+s}.
\end{equation}
We assume this step is irreversible, there is no fragmentation of
the glycine aggregate. Below, we simplify this so that only dimers
are formed ($r=1,s=1$)\cite{Lahav2005b}.

\textsl{Nucleation of \textit{oriented} glycine crystals at the
air/water interface in presence of hydrophobic $LD$-amino acids.}
The hydrophobic amino acid at the interface acts as a nucleator or
seed (see also Scheme 5 in \cite{Lahav2005}):
\begin{eqnarray}\label{gnucleate}
A_r  +  L_1 &\stackrel{\mu_r}{\longrightarrow}& \{L_1X_r\},\nonumber\\
A_r  +  D_1 &\stackrel{\mu_r}{\longrightarrow}& \{D_1Y_r\}.
\end{eqnarray}
Here and henceforth, the brackets denote the solid phase. This
process occurs at the rate $\mu_r$. The two enantiotopic crystal
faces exposed to the solution are denoted by $X_r$ and $Y_r$,
respectively. Realistically, a single leucine or valine molecule
cannot serve as a nucleus for the glycine crystallization, instead,
several dozens arranged in self-assembly are usually required.
However, for the sake of simplicity (and limitations on
computational time), we assume that one hydrophobic amino acid
monomer is sufficient to trigger the nucleation. Otherwise, we would
have to additionally model the formation of the chiral template
clusters formed by several hydrophobic amino acids at the interface
\cite{Lahav1987,Lahav1989}. This template, whatever its actual size,
is incorporated into the growing glycine crystal, but in the face
exposed to air. We note that HPLC analysis of the crystals indicates
that only minute amounts of the orienting hydrophobic aminos acids
are occluded through the upward pointing face\cite{Lahav1984}, so
our single monomer ``template" assumption is not unreasonable and
results in a welcome mathematical simplification. Varying the size
of the template can only have quantitative but not qualitative
effects in so far as mirror symmetry breaking is concerned. Below,
we will assume a transition regime, that is, only dimer glycine
clusters $r=2$ get nucleated in this way to form oriented host
seeds.

\textsl{Enantioselective occlusion of the amino acid monomers from
solution.} Incorporation of the guest molecules (amino acids) into
mixed host crystal leading to formation of homochiral mixed crystal.
As dictated by the actual experimental results \cite{Lahav1984}, we
assume that the $X$ face exposed to solution occludes only the
$D$-amino acids, whereas the $Y$ face occludes only the $L$-amino
acids:
\begin{eqnarray}\label{occlusion}
\{L_1X_r D_n\}  +  D_1 &\stackrel{\xi_{r,n}}{\longrightarrow}& \{L_1X_{r}D_{n+1}\}, \nonumber\\
\{D_1Y_r L_n\}  +  L_1 &\stackrel{\xi_{r,n}}{\longrightarrow}&
\{D_1Y_{r}L_{n+1}\}.
\end{eqnarray}
The corresponding rate $\xi_{r,n}$ could depend on both the size $r$
of the glycine host as well as on the number $n$ of previously
occluded enantiomers.  We assume the oriented host glycine crystal
occludes one amino-acid monomer at a time. Crystallographic models
suggest that single molecules, not clusters, are incorporated one at
a time into the growing crystal.  The notation in $\{...\}$ is such
that reading from left to right: the orienting hydrophobic amino
acid attached to the face exposed to air, the enantiotopic crystal
face exposed to solution (composed of $r$-glycine monomers) and the
number $n$ of occluded amino acids from solution. There is no
structural or sequence information implied. Note this step implies
that the percent $ee$ of occluded enantiomers per host crystal will
be $100\%$, since each enantiotopic crystal face occludes
enantioselectively.  This is fully justified by the actual
experimental results, see the fourth column of Table I in
reference\cite{Lahav1984}.

\textsl{Growth of the oriented host glycine crystal.} Growth of the
enantiotopic face exposed to solution by take-up of the achiral
glycine monomers from the solution:
\begin{eqnarray}\label{gnucleate}
\{L_1X_r D_n\}  +  A_1 &\stackrel{\alpha_{r,n}}{\longrightarrow}& \{L_1X_{r+1}D_n\}, \nonumber\\
\{D_1Y_r L_n\}  +  A_1 &\stackrel{\alpha_{r,n}}{\longrightarrow}&
\{D_1Y_{r+1}L_n\}.
\end{eqnarray}
The newly acquired monomers adopt the same orientation as their
host-face. Assumption: take-up involves one glycine monomer at a
time. Here the rate $\alpha_{r,n}$ could in principle depend on $r$
and $n$.

\begin{figure}[h]
\centering
  \includegraphics[width=0.50\textwidth]{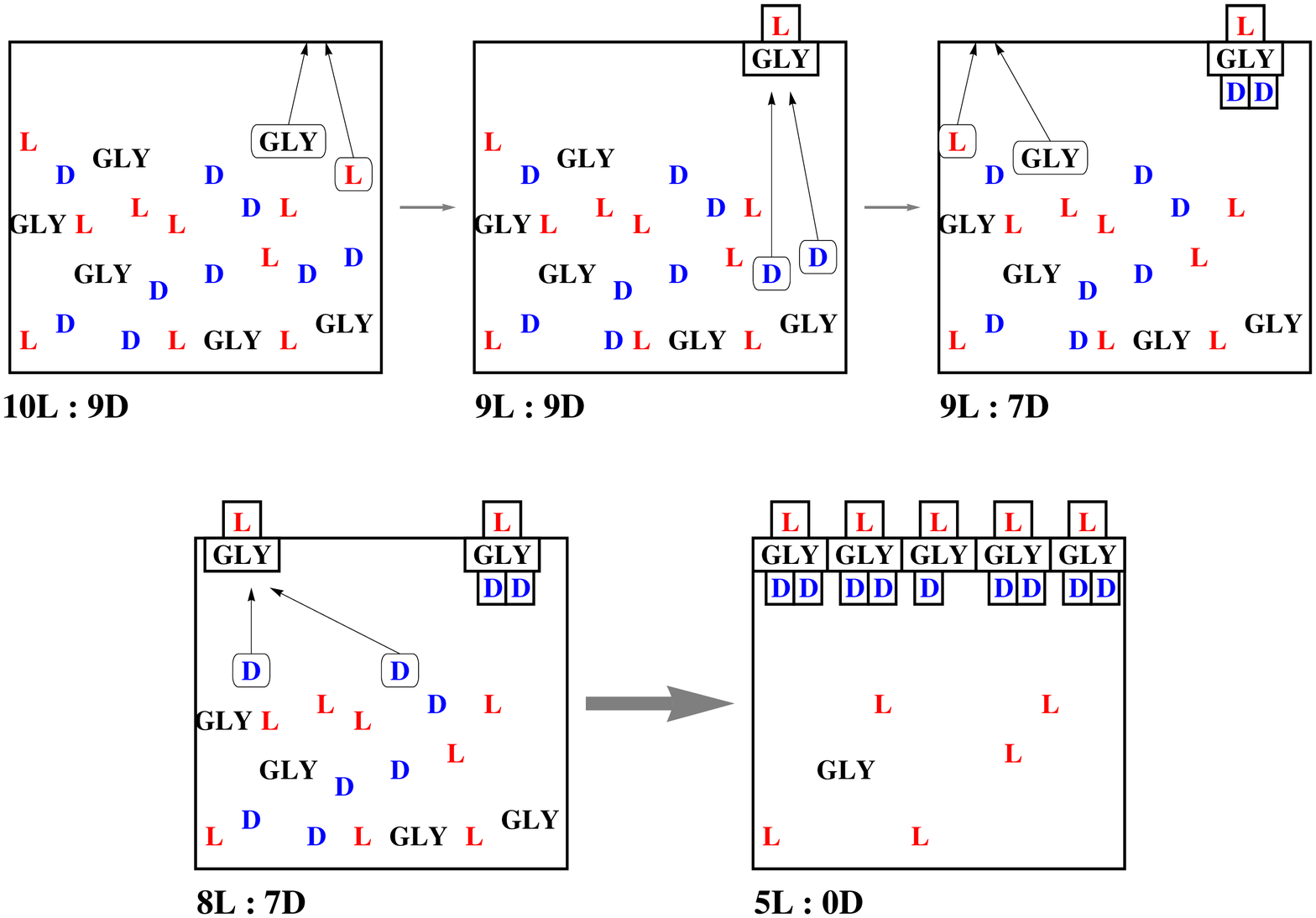}
  \caption{Simplified schematic representation of our model for the generation and amplification
  of optical activity by enantioselective occlusion of amino acids by floating glycine crystals oriented at the
  air/water interface by
  hydrophobic additives. The system starts from an initial enantiomeric excess $L>D$ in solution.
  The horizontal arrows indicate the temporal sequence of events. }
  \label{Whyitworks}
\end{figure}

The above steps give a specific articulation of the Scheme
represented in Figure \ref{SchemeII}. The sequence of pictures in
Figure \ref{Whyitworks} illustrates how the the hydrophobic
orienting effect plus the enantioselective occlusion work in tandem
to yield resolution and optical activity. For illustrative purposes
only, we consider a model in which one hydrophobic amino acid is
sufficient to orient crystals at the air/water interface (see the
above remarks) and that the crystals can occlude up to two amino
acids from the solution. The events are ordered from left to right
as indicated by the arrows. Start from a closed system containing a
pool of amino acid enantiomers in an initial ratio of $L:D=10:9$
plus an initial concentration of glycine aggregates, Gly. By chance,
a glycine aggregate nucleates with its $(010)$ exposed to the
solution, this step uses up one hydrophobic $L$ from the solution so
now we have $L:D=9:9$. The crystal starts occluding the $D$ monomers
from a solution and it therefore becomes enriched in the $L$
enantiomer: $L:D=9:7$. It is more probable that the \textit{next}
crystal to be nucleated at the interface will have the same
orientation as the Adam crystal. Another $L$ is needed to orient the
crystal and so two more $D$ monomers get occluded. The solution is
further enriched in the $L$ enantiomer. In the course of time, this
yields a cascade mechanism and results in a final configuration with
a fully oriented glycine crust with a resolved amino acids for each
individual crystal and an optically active solution rich in the
$L$-enantiomer. The finite surface area gets covered by nucleated
glycine crystals, and the reactions at the interface cease when
either the maximum number of $D$'s per crystal are occluded or when
they are depleted from the solution.

Note in passing: there no explicit \textit{spatial} or coordinate
dependence in this model. We do not explicitly distinguish the
interface surface from the bulk solution. We can think of this as a
homogeneous two-dimensional model, namely, the two-dimensional
air/water interface, with the role of the three-dimensional bulk
solution (below the interface) as providing the interface with the
new glycine and amino acid molecular building blocks. Also, there is
no fragmentation; once formed the crystals do not break up into
smaller pieces, the guest molecules stay with their hosts. The real
experimental situation is three-dimensional, being composed of an
interface plus bulk solution. It is clear that the molecules in
solution must diffuse in order to reach the interface. Even the
crust formed at the interface is not homogeneous, but is composed of
the fusion of many plate-like crystals \cite{Lahav1988}.  We ignore
size distribution of the crystals grown at the interface. But for
the purposes of obtaining a simple model that leads to
experimentally established resolution, these details will not
matter. When we treat the complete truncated model below, we will
however implement a way to effectively account for the
\textit{finite} surface area available for the oriented glycine
crystals.

\section{Mathematical model}

We transcribe the above reaction processes into the corresponding
differential rate equations. Introduce concentration variables
$a_r(t) = [A_r]$, $L_1(t) = [L_1]$, $D_1(t) = [D_1]$ and for the two
host-plus-guest crystal orientations exposed towards the solution
$f_{r,n}(t) =\{L_1X_rD_n\}$ and $\bar{f}_{r,n}(t) =\{D_1Y_rL_n\}$,
respectively. We consider separately the kinetic equations for
$f_{r,0}$, $f_{2,n}$, and then $f_{r,n}$ and similarly for the other
enantiotopic face orientation. That is, we can treat the growth of
pure host crystal, growth of guest on minimal host substrate and
then the general \textit{mixed} growth of both host and guest.
Remember $f_{r,k} = f_{host,guest}$, so there must be a minimum host
size $r$ to accommodate the occluded guests.

Applying law of mass action we obtain the following kinetic
equations:
\begin{eqnarray}
\label{glycine_mon}
\frac{d a_1}{dt} &=& -\mu_1 a_1(L_1 + D_1)
-2\delta_{11}a_1^2
-\sum_{k=2}^{\infty}\delta_{1,k}a_1 a_k \nonumber\\
&&- a_1\sum_r \sum_{n=0}\alpha_{r,n}(f_{r,n}+{\bar f}_{r,n}),\\
\label{glycine_r} \frac{d a_r}{dt} &=& -\mu_r a_r(L_1 + D_1) +
\frac{1}{2}\sum_{k=1}^{r-1} \delta_{k,r-k}a_k a_{r-k}\nonumber\\
&&-
\sum_{k=1}^{\infty} \delta_{r,k}a_r a_k, \,\,(r\geq 2)\\
\label{L1} \frac{dL_1}{dt} &=& -L_1 \sum_{r} \mu_r a_r -L_1\sum_r
\sum_{n=0} \xi_{r,n}{\bar f}_{r,n},\\
\label{D1} \frac{dD_1}{dt} &=& -D_1 \sum_{r} \mu_r a_r -D_1\sum_r
\sum_{n=0}
\xi_{r,n}f_{r,n},\\
\label{purefhost} \frac{d f_{r,0}}{dt} &=& \mu_r L_1 a_r
-\xi_{r,0}f_{r,0}D_1 \nonumber\\
&&+ a_1(\alpha_{r-1,0}f_{r-1,0} - \alpha_{r,0}f_{r,0}),\\
\label{purebfhost} \frac{d {\bar f}_{r,0}}{dt} &=& \mu_r D_1 a_r
-\xi_{r,0}{\bar f}_{r,0}L_1 \nonumber\\
&&+a_1(\alpha_{r-1,0} {\bar f}_{r-1,0} - \alpha_{r,0}{\bar
f}_{r,0}),\\
\label{fhostguest} \frac{d f_{r,n}}{dt} &=&
D_1(\xi_{r,n-1}f_{r,n-1} -\xi_{r,n}f_{r,n} )\nonumber\\
&&+a_1 (\alpha_{r-1,n}f_{r-1,n} - \alpha_{r,n}f_{r,n}),(n\geq 1) \\
\label{bfhostguest} \frac{d {\bar f}_{r,n}}{dt} &=&
L_1(\xi_{r,n-1}{\bar f}_{r,n-1} -\xi_{r,n}{\bar f}_{r,n})\nonumber\\
&&+ a_1(\alpha_{r-1,n}{\bar f}_{r-1,n} - \alpha_{r,n}{\bar
f}_{r,n}),(n\geq 1)\nonumber.\\
\end{eqnarray}

For a closed system, we have the constant density constraint:
\begin{equation}
L_1 + D_1 + \sum_r^{\infty} ra_r +
\sum_r^{\infty}\sum_n^{\infty}(r+n+1)(f_{r,n} + {\bar f}_{r,n}) =
const.
\end{equation}
The main problem with such a model is the large number of parameters
that have been introduced: namely $\delta_{r,n},\mu_r,\alpha_{r,n}$
and $\xi_{r,n}$. It is thus clear we need to make several
simplifications already at this stage.

\subsection{Simplifications of the model}

We assume that only the coalescence of two glycine monomers $A_1$ to
form an $A_2$ dimer needs to be retained, and only $A_2$ dimer
clusters are used to form new oriented host crystals when they
encounter a hydrophobic amino acid monomer. We also assume that all
rates are independent of cluster size, so that
\begin{eqnarray}
\delta_{1,1} &=& \delta, \quad\qquad\qquad \delta_{r,k} = 0, (otherwise) \\
\mu_2 &=& \mu, \quad\qquad\qquad \mu_r = 0, (r=1 \& r\geq 3) \\
\alpha_{r,n} &=&  \alpha, (r\geq 2),\qquad \alpha_{1,n}=0 \\
\xi_{r,n} &=& \xi , (r\geq 2),\qquad \xi_{1,n}=0.
\end{eqnarray}
Then the above kinetic equations take the following form:
\begin{eqnarray}
\frac{da_1}{dt} &=& -2\delta a_1^2 - \alpha
a_1\sum_{r=2}\sum_{n=0}(f_{r,n} + {\bar f}_{r,n}),\\
\frac{da_2}{dt} &=& \delta a_1^2 -\mu a_2(L_1 + D_1),\\
\frac{dL_1}{dt} &=& -\mu L_1 a_2 - \xi L_1 \sum_{r=2}\sum_{n=0}
{\bar f}_{r,n},\\
\frac{dD_1}{dt} &=& -\mu D_1 a_2 - \xi D_1 \sum_{r=2}\sum_{n=0}
f_{r,n},\\
\frac{d f_{2,0}}{dt} &=& \mu L_1 a_2 - (\xi D_1  + \alpha a_1)
f_{2,0},\\
\frac{d {\bar f}_{2,0}}{dt} &=& \mu D_1 a_2 - (\xi L_1  + \alpha
a_1)
{\bar f}_{2,0},\\
\frac{d f_{r,0}}{dt} &=& -\xi D_1 f_{r,0} + \alpha a_1(f_{r-1,0} -
f_{r,0}), \quad (r\geq 3)\\
\frac{d {\bar f}_{r,0}}{dt} &=& -\xi L_1 {\bar f}_{r,0} + \alpha
a_1({\bar f}_{r-1,0} -
{\bar f}_{r,0}), \quad (r\geq 3)\\
\frac{d f_{2,n}}{dt} &=&  \xi D_1(f_{2,n-1}-f_{2,n}) - \alpha a_1 f_{2,n},\quad (n\geq 1)\\
\frac{d {\bar f}_{2,n}}{dt} &=&  \xi L_1({\bar f}_{2,n-1}- {\bar f}_{2,n}) - \alpha a_1 {\bar f}_{2,n},\quad (n\geq 1)\\
\frac{d f_{r,n}}{dt} &=&  \xi D_1(f_{r,n-1} - f_{r,n} ) + \alpha a_1 (f_{r-1,n} - f_{r,n}),\nonumber \\
&& \qquad (r\geq 3, n\geq 1)\\
\frac{d {\bar f}_{r,n}}{dt} &=& \xi L_1({\bar f}_{r,n-1} - {\bar
f}_{r,n}) + \alpha a_1({\bar f}_{r-1,n} - {\bar
f}_{r,n}),\nonumber \\
&& \qquad (r\geq 3, n\geq 1).
\end{eqnarray}
As an important check on internal consistency, we verify explicitly
that mass conservation holds.  To this end, we consider the
following change of variables
\begin{eqnarray}
P &=& \sum_{r=3}^{\infty} f_{r,0}, \quad M_2 = \sum_{n=1}^{\infty}
f_{2,n}, \quad
M = \sum_{r=3}^{\infty}\sum_{n=1}^{\infty} f_{r,n},\\
\bar P &=& \sum_{r=3}^{\infty} {\bar f}_{r,0}, \quad \bar M_2 =
\sum_{n=1}^{\infty} {\bar f}_{2,n}, \quad \bar M =
\sum_{r=3}^{\infty}\sum_{n=1}^{\infty} {\bar f}_{r,n}.
\end{eqnarray}
Then the above infinite collection of equations can be written in
terms a mathematically closed set of just twelve equations:
\begin{eqnarray}
\frac{d a_1}{dt} &=& -2\delta a_1^2 \nonumber\\
&&- \alpha a_1\Big(f_{2,0} + M_2 +
P + M + \bar f_{2,0} + \bar P + \bar M + \bar M_2\Big),\nonumber\\
\\
\frac{d a_2}{dt} &=& \delta a_1^2 - \mu a_2(L_1 + D_1),\\
\frac{d L_1}{dt} &=& -\mu L_1 a_2 -\xi L_1\Big( \bar f_{2,0} + \bar
P + \bar M + \bar M_2\Big),\\
\frac{d D_1}{dt} &=& -\mu D_1 a_2 -\xi D_1\Big( f_{2,0} + M_2 +
P + M \Big),\\
\frac{d f_{2,0}}{dt} &=& \mu L_1 a_2 - (\xi D_1  + \alpha a_1)
f_{2,0},\\
\frac{d {\bar f}_{2,0}}{dt} &=& \mu D_1 a_2 - (\xi L_1  + \alpha
a_1)
{\bar f}_{2,0},\\
\frac{d P}{dt} &=& -\xi D_1 P + \alpha a_1 f_{2,0}, \\
\frac{d \bar P}{dt} &=& -\xi L_1 \bar P + \alpha a_1 \bar f_{2,0},\\
\frac{d M_2}{dt} &=& \xi D_1 f_{2,0} - \alpha a_1 M_2, \\
\frac{d \bar M_2}{dt} &=& \xi L_1 \bar f_{2,0} - \alpha a_1 \bar
M_2, \\
\frac{d M}{dt} &=& \xi D_1 P + \alpha a_1 M_2,\\
\frac{d \bar M}{dt} &=& \xi L_1 \bar P + \alpha a_1 \bar M_2.
\end{eqnarray}
We need to have kinetic equations for the corresponding densities.
The following definitions will accomplish this:
\begin{eqnarray}
&&\rho_P = \sum_{r=3}^{\infty} (r+1)f_{r,0},\\
&&\rho_{M2} = \sum_{n=1}^{\infty} (n+3)f_{2,n},\\
&&\rho_{M} = \sum_{r=3}^{\infty} \sum_{n=1}^{\infty} (r+n+1)f_{r,n}.
\end{eqnarray}
Then the kinetic equations for the densities follow:
\begin{eqnarray}
\dot \rho_P &=& -\xi D_1 \rho_P + \alpha a_1(4f_{2,0} + P),\\
\dot \rho_{M2} &=& 4\xi D_1 f_{2,0} + \xi D_1 M_2 - \alpha
a_1\rho_{M2}, \\
\dot \rho_M &=& \xi D_1\Big(P + \rho_P + M \Big) + \alpha a_1 \Big(
\rho_{M2} + M_2 + M \Big).
\end{eqnarray}
There are of course analogous equations for the $Z_2$ or chiral
partners $\bar \rho_{P}, \bar \rho_{M2}$ and $\bar \rho_M$. To
express the mass conservation in terms of these densities, define
\begin{eqnarray}
\rho &=& 3f_{2,0} + \rho_{M2} + \rho_{P} + \rho_{M}, \\
\bar \rho &=& 3\bar f_{2,0} + \bar \rho_{M2} + \bar \rho_{P} + \bar
\rho_{M},
\end{eqnarray}
then the following constraint holds
\begin{equation}
a_1 + L_1 + D_1 + 2a_2 + \rho + \bar \rho = const.
\end{equation}
This proves that total mass of the reactants is conserved in our
scheme.

\subsection{Linear Stability}
It is useful at this stage to check if the model is actually capable
of mirror symmetry breaking. Consider the following variables where
$H$ denotes pure host crystals, and $M$ the mixed host plus guest
crystals:
\begin{eqnarray}
H &=& f_{2,0} + P,\\
\bar H &=& \bar f_{2,0} + \bar P,\\
M &=& M_2 + M, \\
\bar M &=& \bar M_2 + \bar M.
\end{eqnarray}
Then we can further reduce the system from twelve to only eight
differential equations:
\begin{eqnarray}\label{trunc1}
\frac{d a_1}{dt} &=& -2\delta a_1^2 - \alpha a_1\Big(H + \bar H + M + \bar M \Big),\\
\frac{d a_2}{dt} &=& \delta a_1^2 - \mu a_2(L_1 + D_1),\\
\frac{d L_1}{dt} &=& -\mu L_1 a_2 -\xi L_1\Big( \bar H + \bar M\Big),\\
\frac{d D_1}{dt} &=& -\mu D_1 a_2 -\xi D_1\Big( H + M\Big),\\
\frac{d H}{dt} &=& \mu L_1 a_2 - \xi D_1 H,\\
\frac{d \bar H}{dt} &=& \mu D_1 a_2 - \xi L_1 \bar H,\\
\frac{d M}{dt} &=& \xi D_1 H,\\ \label{trunc8}
\frac{d \bar M}{dt}
&=& \xi L_1 \bar H.
\end{eqnarray}
We can define three types of chiral polarizations $\eta, \theta,
\phi$. Investigate mirror symmetry breaking by the change of
variables:
\begin{eqnarray}
\chi &=& L_1 + D_1,\, w = H + \bar H, \, z = M + \bar M,\\
\eta &=& \frac{L_1 - D_1}{\chi}, \, \theta = \frac{H- \bar H}{w}, \,
\phi = \frac{M - \bar M }{z}.
\end{eqnarray}
The transformed equations Eqs.(\ref{trunc1}-\ref{trunc8}) read as
follows:
\begin{eqnarray}
\frac{d a_1}{dt} &=& -2\delta a_1^2 - \alpha a_1 ( w + z ),\\
\frac{d a_2}{dt} &=& \delta a_1^2 - \mu a_2 \chi,\\
\frac{d \chi}{dt} &=& -\mu a_2 \chi -\frac{\xi}{2}\chi\Big(w(1-\eta \theta) + z(1 - \eta \phi)\Big),\\
\frac{d \eta}{dt} &=& \frac{\xi}{2}(1 - \eta^2)(w\theta + z\phi),\\
\frac{d w}{dt} &=& \mu a_2 \chi - \frac{\xi \chi w}{2}(1 - \eta \theta),\\
\frac{d \theta}{dt} &=& \frac{\mu a_2 \chi}{w}(\eta - \theta) + \frac{\xi \chi}{2}\eta(1 - \theta^2),\\
\frac{d z}{dt} &=& \frac{\xi \chi w}{2}(1 - \eta \theta),\\
\frac{d \phi}{dt} &=& \frac{\xi \chi w}{2z}(\theta - \eta) -
\frac{\xi \chi w}{2z}\phi(1 - \eta \theta).
\end{eqnarray}
Consider the stability of the symmetric solution $\eta = \theta =
\phi = 0$. Linearize the system of the three corresponding rate
equations and determine stability of the solution from the Jacobian
matrix:
\begin{equation}
\left(
  \begin{array}{ccc}
    0 & \frac{\xi}{2}w & \frac{\xi z}{2} \\
    \frac{\mu a_2\chi}{w} + \frac{\xi \chi}{2} & -\frac{\mu a_2 \chi}{w} & 0 \\
    -\frac{\xi}{2z}\chi w & \frac{\xi}{2z}\chi w & -\frac{\xi}{2z}\chi w \\
  \end{array}
\right).
\end{equation}
Expansion along the minors yields the determinant $\det$ which is
also given by the product of the three eigenvalues $\lambda_i$:
\begin{equation}\label{det}
\det = + \frac{\xi^2}{4z}\chi^2 w^2\Big(\frac{\mu a_2}{w} +
\frac{\xi}{2} \Big) + \frac{\xi^3 \chi^2 w}{8}= \prod_{i=1}^3
\lambda_i
> 0.
\end{equation}
This result immediately tells us that the symmetric or racemic state
is \textit{unstable}. To confirm this, first consider the case where
all three eigenvalues $\lambda_i$ are real. Then the only way for
their product to be positive is if the individual eigenvalues have
the algebraic signs $(+,+,+)$, or $(+,-,-)$ (the order is not
important): the point is, either one eigenvalue or else all three
must be positive, and thus in both cases the \textit{symmetric state
is unstable}. So chiral symmetry will be broken in this infinite but
closed model. Note that the remaining sign possibilities $(-,-,-)$
and $(+,+,-)$ (again, the order is not important) would yield
instead a negative determinant $\det<0$, and could imply stable or
unstable, respectively. This latter situation is inconclusive
without further analysis. If on the other hand $\det$ has complex
eigenvalues, these always occur in complex conjugate pairs, and
without loss of generality we may suppose that $\lambda_3 =
{\lambda_2}^*$. Then $\det = \lambda_1 |\lambda_2|^2
> 0$ if and only if $\lambda_1 >0$, and again we conclude that the
symmetric state is unstable. The expression for $\det$ depends on
the rate of enantioselective occlusion $\xi$ and the rate for
hydrophobic orientation of host crystals $\mu$ and suggests that the
former is more important than the latter for driving the symmetry
breaking: this is so because we can set $\mu = 0$ in Eq. (\ref{det})
and $\det>0$ remains positive, so the racemic state remains
unstable. But if we set $\xi = 0$, then $\det$ vanishes identically
and the result is inconclusive at this lowest order. On the other
hand, we note that neither the rate of glycine up-take $\alpha$ nor
the rate of glycine cluster formation $\delta$ appear explicitly in
the determinant, suggesting that these processes in and of
themselves are \textit{not} the decisive factors for symmetry
breaking. These expectations are in accord with the processes
depicted in the scheme represented in Fig. \ref{SchemeII}: the
crucial amplification cycle involves the orientation and occlusion
steps only. The processes of glycine aggregation in solution
($\delta $) and growth by glycine up-take ($\alpha$) are
\textit{prior} events that are not involved in this cycle. Thus, at
this stage, we may be confident that our model will qualitatively
capture the main symmetry breaking aspects of the actual experiment,
and that the bare minimal ingredients required for this are crystal
orientation and enantioselective occlusion.

The above model allows unlimited growth of the mixed host plus guest
crystals. A severe \textit{truncation} yielding a minimal model
consists of an $r=2$ glycine host with at most one occluded amino
acid guest $n=1$ per host. The determinant in this case is given by
Eq.(\ref{det}) after deleting the $\xi^3 \chi^2 w/8$ term. The
linear stability analysis carried out for this truncated model leads
to the same conclusions as above. We thus find that the symmetric
state is unstable in two opposite limiting cases of the underlying
model: for glycine host dimers and one occluded guest as well as for
arbitrarily large glycine host crystals $r < \infty$ accommodating
an unlimited number $n < \infty$ of occluded guests. Thus we may
expect that the symmetric state will be unstable for all
intermediate host/guest truncations of the model. This expectation
is confirmed by the numerical simulations presented below.

\section{Complete truncated model}

The model leads to the unlimited growth of both the glycine host
crystals as well as to an unlimited number of enantiomers that can
be occluded per host crystal. The experiments by contrast yield
finite size mixed glycine plus host crystals with a rather uniform
size distribution and with each host occluding a small percentage of
the available amino acids\cite{Lahav1984,Lahav1988}. We must also
recognize the fundamental physical limitation imposed by the fact
that the air/water interface is of finite area: the experiments are
carried out in a bounded reaction domain which means only a finite
air/water interface is available for the crucial
orientation/amplification processes to take place. We will thus need
to model the effect of a bounded (finite area) interface. Once the
interface is \textit{covered} by floating glycine crystals, no more
hydrophobic amino acids can diffuse up to the surface, the remaining
reactions from this point on can only be the glycine up-take from
the solution and the enantioselective occlusion of the amino acids
in solution. The finite size of the experiment implies that these
latter two processes must be limited as well; we will thus consider
how to truncate both the host growth as well as the number of
occluded guests per host. The reactions will thus stop at some
point: there must be maximum values of both $R$ and $N$ such that $2
\leq r \leq R$ and hosts $0 \leq n \leq N(R)$. Each instantaneous
size $r$ of the host crystal will be allowed to occlude up to
maximum number $n(r)$ of guests. We can account for all of these
limiting features by implementing certain truncations or cut-offs
that we apply to the underlying model. At the same time, we complete
the model by including the kinetic inhibition effect.

\subsection{Hydrophobic and kinetic effects and hydrophobic and hydrophilic additives}

The experiment demonstrates that the glycine crystals at the
interface are oriented via two distinct effects: (i) hydrophobic and
(ii) kinetic. The former is due to the induction by hydrophobic
amino acids while the latter is achieved through the inhibition of
nucleation and growth of the oriented crystals. The hydrophobic
effect is confined to hydrophobic amino acids but the kinetic effect
applies to all $\alpha$-amino acid \textit{additives}, both
hydrophobic and hydrophilic. It is important to bear in mind that
there is hence not a one-to-one correspondence between effects and
additives. For this reason, in the full model below, we introduce
the two kinds of additives and talk about the effect of the
additives rather than the hydrophobic or kinetic effects \textit{per
se}, since the latter are not easily separable. The two effects
however do act in the same direction. The inclusion of both types of
additive leads to only a minor modification and leads to a complete
model. We also implement a consistent truncation where $R$ is
maximum size of the glycine host crystal and $n=n(r)$ is a maximum
number of occluded amino acid guest monomers for any instantaneous
host size $r$.

For completeness, we list the full set of chemical reactions
defining the final model. Formation of glycine crystals/clusters in
solution.
\begin{equation}\label{gcluster}
A_1  + A_1 \stackrel{\delta}{\longrightarrow} A_2.
\end{equation}

Diffusion of amino acids in bulk solution to interface implying
nucleation of \textit{oriented} glycine crystals at the air/water
interface in presence of hydrophobic $LD$-amino acids.
\begin{eqnarray}\label{diffuse}
A_2+L_1 &\stackrel{\mu(f_c)}{\longrightarrow}& \{L_1 X_2\},\nonumber\\
A_2+D_1 &\stackrel{\mu(f_c)}{\longrightarrow}& \{D_1 Y_2\} .
\end{eqnarray}
The effective conversion rate of bulk amino acids to amino acids at
the interface will depend on a critical glycine crystal interface
concentration $f_c$. Beyond this concentration, no more hydrophobic
amino acids will diffuse up to the surface, and consequently no more
fresh glycine crystals can nucleate at the surface. To implement
this we will take
\begin{equation}
\mu(f_c) = \mu_0 \Theta\Big( f_c - f(t) \Big),
\end{equation}
where $f(t)$ is be the instantaneous concentration of glycine
crystals, with and without occluded amino acids, at time $t$, and
$f_c$ is a critical surface concentration which is supposed to
effectively mock the finite interface. Here:
\begin{equation}
\mu(t)= \mu_0\Theta\Big(f_c -
\sum_{r=3}^{R}\sum_{n=0}^{\Gamma}(f_{r, n}(t) + \bar{f}_{r,
n}(t)\Big),
\end{equation}
where the unit step function is defined as
\begin{equation}
\Theta(x) = \left\{
\begin{array}{cc}
   1, & x > 0 \\
   0, & x < 0.
 \end{array}
 \right.
\end{equation}
Once the instantaneous concentration of glycine crystals reaches
$f_c$ then this reaction shuts off, from which point on the
hydrophobic orienting effect ceases to act. This is an effective way
of implementing the finite interface area constraint without adding
complicated spatial dependence to the model.

In keeping with the above remarks and experimental facts, we must
allow for the hydrophobic amino acids to inhibit those glycine
nuclei exposing their enantiotopic faces towards the solution from
further growth.  We also introduce a second species of hydrophilic
amino acids ${\bar L},{\bar D}$, which can only participate in the
kinetic effect.

Inhibition of the fresh glycine seeds by kinetic effect due to the
\textit{hydrophobic} additives.
\begin{eqnarray}\label{occlusion}
\{L_1X_2\}+D_1 &\stackrel{\xi}{\longrightarrow}& \{L_1X_2D_1\}, \nonumber\\
\{D_1Y_2\}+L_1 &\stackrel{\xi}{\longrightarrow}& \{D_1Y_2L_1\}.
\end{eqnarray}
Inhibition of the fresh glycine seeds by kinetic effect due to the
\textit{hydrophilic} additives:
\begin{eqnarray}\label{occlusion}
\{L_1X_2\}+\bar{D}_1 &\stackrel{\beta}{\longrightarrow}& \{L_1X_2\bar{D}_1\}, \nonumber\\
\{D_1Y_2\}+\bar{L}_1 &\stackrel{\beta}{\longrightarrow}&
\{D_1Y_2\bar{L}_1\},
\end{eqnarray}
the total inert product of this inhibited crystal seeds (via either
the hydrophobic or hydrophilic additives) will be denoted as $P(t)$.
An excess of $L>D$ in solution inhibits glycine nuclei exposing
their $(0\bar 1 0)$ towards solution\cite{Lahav2005}, preventing
their further growth, and analogously for $D>L$. We have opted to
model this as a cross-inhibition reaction, in the spirit of the
Frank model, except here we form a small tetramer unit with a
glycine dimer trapped between the hydrophobic templater and the
hydrophilic growth inhibitor. The templater and growth inhibitor
have opposite handedness.  So, if the glycine host crystal occludes
an amino acid having the opposite chirality of the nucleator amino
and located on the opposite enantiotopic face, then the seed glycine
crystal has been inhibited and can not grow further. Otherwise, if
the seed grows by incorporating a glycine monomer, it can escape
this ``bottleneck", and will continue to grow via glycine up-take or
by amino acid occlusion.

The remainder of the reactions are unchanged, except that now we
impose the finite size truncations (bounds on $r$ and on $n$)
remarked above. Growth of the oriented fresh glycine hosts.
\begin{eqnarray}\label{occlusion}
\{L_1X_2\}+A_1 &\stackrel{\alpha}{\longrightarrow}& \{L_1X_3\}, \nonumber\\
\{D_1Y_2\}+A_1 &\stackrel{\alpha}{\longrightarrow}& \{D_1Y_3\}.
\end{eqnarray}

Growth of the oriented host glycine crystal.
\begin{eqnarray}\label{gnucleate}
&&\{L_1X_r D_n\}  +  A_1 \stackrel{\alpha}{\longrightarrow} \{L_1X_{r+1}D_n\}, \nonumber\\
&&\{D_1Y_r L_n\}  +  A_1 \stackrel{\alpha}{\longrightarrow}
\{D_1Y_{r+1}L_n\}, \nonumber\\ &&\qquad(3\leq r \leq R-1, 0\leq n
\leq
\Gamma(r)).\nonumber\\
\end{eqnarray}

Enantioselective occlusion of the amino acid monomers from solution.
\begin{eqnarray}\label{occlusion}
&&\{L_1X_r D_n\}  +  D_1 \stackrel{\xi}{\longrightarrow} \{L_1X_{r}D_{n+1}\}, \nonumber\\
&&\{D_1Y_r L_n\}  +  L_1 \stackrel{\xi}{\longrightarrow}
\{D_1Y_{r}L_{n+1}\}, \nonumber\\ &&\qquad(3\leq r \leq R, 0\leq n
\leq
\Gamma(r)-1).\nonumber\\
\end{eqnarray}
Each host glycine crystal can grow by incorporating achiral glycine
monomers from the solution, so we must impose a maximum number $R$
of glycine monomers forming the host crystal. By the same token, we
impose a maximum number of guest chiral monomers $N$ to be occluded
in the host crystal. The maximum number of guests that a crystal can
occlude depends on the number of host monomers of the crystal, so we
define $\Gamma(r)=Floor[\gamma r]$, where $Floor[\gamma r]$ rounds
the elements of $\gamma r$ to the nearest integer less than or equal
to $\gamma r$ and $0 < \gamma \leq 1$ is a free parameter that
allows us to vary the percentage of occluded amino acids.

With these truncations and limits and for both types of additives,
the final set of kinetic equations reads as follows:
\begin{eqnarray}\label{firsteq}
\frac{da_{1}}{dt}&=&-2\delta a_{1}^{2}\nonumber\\
&&-\alpha a_{1}(f_{2,0}+\bar{f}_{2,0}+ \sum_{r=3}^{R-1}\sum_{n=0}^{\Gamma(r)}(f_{r,n}+\bar{f}_{r,n})), \\
\frac{da_{2}}{dt}&=&\delta a_{1}^{2}-\mu a_{2}(L_{1}+D_{1}), \\
\frac{dL_{1}}{dt}&=&-\mu L_{1} a_{2}-\xi L_{1}(\bar{f}_{2,0}+\sum_{r=3}^{R}\sum_{n=0}^{\Gamma(r)-1}\bar{f}_{r,n}),\\
\frac{dD_{1}}{dt}&=&-\mu D_{1} a_{2}-\xi D_{1}(f_{2,0}+\sum_{r=3}^{R}\sum_{n=0}^{\Gamma(r)-1}f_{r,n}),\\
\frac{d\bar{L}_1}{dt}&=&-\beta \bar{f}_{2,0} \bar{L}_1,\\
\frac{d\bar{D}_1}{dt}&=&-\beta {f}_{2,0} \bar{R}_1,\\
\frac{dP}{dt}&=&\xi (D_1 f_{2,0}+L_1 \bar{f}_{2,0})+\beta (\bar{D}_1 f_{2,0}+\bar{L}_1 \bar{f}_{2,0}),\\
\frac{df_{r,n}}{dt}&=&A(r,n)\mu L_{1} a_{2}+B(r,n)\xi
D_{1}f_{r,n-1}-C(r,n)\xi D_{1}f_{r,n}\nonumber\\
&&+D(r,n)\alpha
a_{1}f_{r-1,n}-E(r,n)\alpha a_{1}f_{r,n}-F(r,n)\beta \bar{D} f_{r,n},\nonumber\\
&&\qquad\qquad\qquad(2\leq r\leq R,0\leq n\leq \Gamma(r))\\
\frac{d\bar{f}_{r,n}}{dt}&=&A(r,n)\mu D_{1} a_{2}+B(r,n)\xi
L_{1}\bar{f}_{r,n-1}-C(r,n)\xi L_{1}\bar{f}_{r,n}\nonumber\\
\label{lasteq} &&+D(r,n)\alpha
a_{1}\bar{f}_{r-1,n}-E(r,n)\alpha a_{1}\bar{f}_{r,n}-F(r,n)\beta \bar{L} \bar{f}_{r,n} \nonumber\\
&&\qquad\qquad\qquad(2\leq r\leq R,0\leq n\leq \Gamma(r)).\\
\nonumber\end{eqnarray}

For a more streamlined presentation as well as for simulation
purposes we prefer to write an single differential equation for the
$f_{r,n}$ and another one for the $\bar{f}_{r,n}$, whose individual
terms are switched on or off depending on the values of $r,n$. To
this end, we have defined the following \emph{switch} functions
$A,B,C,D,E$ and $F$:
\begin{eqnarray}
&&A(r,n)=1 \quad {\rm if} \quad r=2\&n=0\nonumber\\
&&B(r,n)=1 \quad {\rm if} \quad r\geq3\&n\geq1\nonumber\\
&&C(r,n)=1 \quad {\rm if} \quad (r=2\&n=0)||\\
&&\qquad\qquad\qquad\quad(r\geq3\quad\&\quad n\leq\Gamma(r)-1)\nonumber\\
&&D(r,n)=1 \quad {\rm if} \quad (r\geq3\&n\leq\Gamma(r-1))\nonumber\\
&&E(r,n)=1 \quad {\rm if} \quad (r=2\&n=0)||\\
&&\qquad\qquad\qquad\quad(3\leq r\leq R-1)\nonumber\\
&&F(r,n)=1 \quad {\rm if} \quad r=2\&n=0\nonumber\\
\end{eqnarray}

The above set of kinetic equations satisfy the constant density
constraint which we monitor and verify in all the numerical
simulations:
\begin{eqnarray}\label{constraint}
&&L_{1}+D_{1}+ \bar L_1 + \bar D_1 + 4P \nonumber \\
&+&
a_{1}+2a_{2}+\sum_{r=2}^{R}\sum_{n=0}^{\Gamma(r)}(r+n+1)(f_{r,n}+\bar{f}_{r,n})=const.\nonumber\\
\end{eqnarray}

\section{Numerical Results}

We are interested in testing out the effectiveness of the proposed
orientation and amplification cycle (Figure \ref{SchemeII}) as this
is actually represented in the model. At the same time, we can also
control the amounts of each type of additive and assess the relative
importance of the two orienting effects, namely hydrophobic and
kinetic. The results are quantified and presented in terms of a
number of experimentally relevant chiral measures. The percent
enantiomeric excess of the hydrophobic additives in solution is
\begin{equation}\label{eesoln}
ee(\%) = \frac{[L_1]-[D_1]}{[L_1]+[D_1]}\times 100.
\end{equation}
When hydrophilic amino acids are added, a corresponding percent
enantiomeric excess can also be defined:
\begin{equation}\label{eehsoln}
ee_{h}(\%) =
\frac{[\bar{L}_1]-[\bar{D}_1]}{[\bar{L}_1]+[\bar{D}_1]}\times 100.
\end{equation}
We consider the percent excess of the occluded amino acids via
\begin{equation}
ee_{occlude}(\%) = \frac{L_{occluded} - D_{occluded}}{L_{occluded} +
D_{occluded}}\times 100.
\end{equation}
In keeping with the experimental reports, this counts the amino
acids occluded in the host crystal enantiotopic faces exposed to the
solution.  An important aspect of the experiment is the degree of
crystal orientation at the interface. The following parameter
measures the orientation degree of the crystals at the interface:
\begin{equation}\label{eesoln}
od(\%) = \frac{f-\bar{f}}{f+\bar{f}}\times 100.
\end{equation}
and gives us a direct measure of the orientation. Here $f$ and
$\bar{f}$ are the total amount of crystal of each type (i.e.,
pyramids and plates\cite{Lahav1988}) at the surface (omitting the
floating inhibited seeds):
\begin{equation}
f=\sum_{r=3}^{R}\sum_{n=0}^{\Gamma(r)} f_{r,n} \qquad
\bar{f}=\sum_{r=3}^{R}\sum_{n=0}^{\Gamma(r)} \bar{f}_{r,n} \nonumber
\end{equation}
The differential rate equations Eqs. (\ref{firsteq}-\ref{lasteq})
were numerically integrated using the version 7 Mathematica program
package. The results were monitored to verify that the total system
mass Eq. (\ref{constraint}) remained constant in time. We organize
and present our main results in terms of the types of additives
employed, that is,  hydrophobic and/or hydrophilic amino acids.

\subsection{Hydrophobic additives}

To simulate the following cases, only hydrophobic additives have
been added to the saturated glycine solution. So we maintain zero
concentration of the hydrophilic additives $[\bar L_1]_0 = [\bar
D_1]_0= 0$ throughout the following sequence of simulations. All the
numerical results have been obtained for the values $R=50$,
$\gamma=0.1$, $\alpha=10^{-3}$, $\delta=10^{-6}$, $\xi=1$,
$\mu_0=10^{-6}$ and $f_c = 0.01 M$. We provide some brief rationale
for these selected values. The experiment reports that only a small
fraction of the $\alpha$-amino acids $(0.02-0.2\%)$ is subsequently
occluded into the bulk of the growing glycine crystals. We therefore
try to maximize the host crystal size $R$ so as to be able to
achieve small fractions of occluded guests. Computational
limitations (memory and time) forced us to compromise and we thus
choose R=50 and $\gamma=0.1$ (for the desired values
$\gamma=0.0002-0.002$, the corresponding $R$ is so great as to
exceed available computational resources). Next, we aimed to
reproduce as closely as possible the reported details concerning the
growth of crystals at the air/solution interface. The first step was
to find reactions rates satisfying these conditions, as is shown in
our Table 1. Since we do not have spatial information (and thus can
not distinguish between pyramids and plates), we tried to find the
reaction rates (at least the relation between them) reproducing this
behaviour (supposing that pure platforms were the pyramids and the
product of the inhibition yields the plates). This strategy led us
to employ the values of $\alpha, \delta, \xi$  and $\mu$ given
above.

We have used the same glycine concentrations as reported in the
experiments\cite{Lahav1988}. Thus, in a typical experiment 10 g of
glycine ($m_{gly}=10 g$) were dissolved in 30 ml ($Vol=0.03 l$) of
double distilled water, where its molar mass $Mm_{gly}=75.07 g\cdot
mol^{-1}$. Following this, the initial glycine concentration should
therefore be $[a_1]_{0} = m_{gly}/(Mm_{gly}\times Vol)=4.44 \, mol
\cdot l^{-1}$ , and this is the value of initial glycine monomer
concentration used in all the simulations reported here. For the
glycine dimers we initially set $[a_2]_0=0$, and zero initial
concentrations of the oriented glycine crystals at the interface:
$(f_{r,n})_0=0$ and $(\bar f_{r,n})_0 = 0$, respectively. Finally,
from the initial enantiomeric excess $ee_0$ and the relative weight
of additive to glycine, $(w/w_{gly})$, we can deduce the initial
concentrations $[L]_0$ and $[D]_0$ straightforwardly from the pair
of Eqs. (\ref{concentrations}).
\begin{eqnarray}\label{concentrations}
&&[L]_{0} = \frac{(w/w_{gly}) m_{gly} (1+ee_0)}{2 Mm_{gly} \times Vol},\nonumber\\
&&[D]_{0} = \frac{(w/w_{gly}) m_{gly} (1-ee_0)}{2 Mm_{gly} \times
Vol}.
\end{eqnarray}
We ignore the molar mass differences $Mm$ between glycine and the
$\alpha$-amino acid molar masses($Mm(Ala)=89.09 g mol^{-1}$,
$Mm(Leu)=131.17 g mol^{-1}$, $Mm(Val)=117.15 g mol^{-1}$, $Mm(Val)=
105.09 g mol^{-1}$).

\subsubsection{Racemic mixtures of hydrophobic additives}.
In this case, the composition of the complete solution is obtained
by adding a racemic hydrophobic additive to the supersaturated
solution of glycine. This means we start with zero initial excess of
chiral hydrophobic monomers: $ee_{0}=0$. We want to observe the
effect of varying the relative mass of the chiral monomers with
respect to the amount of glycine ($w/w_{gly}$). The numerical
results obtained in these simulations are shown in Table
\ref{rachydrophobic}. For each relative initial concentration of
additives ($w/w_{gly}$) we evaluate the total concentration $f +
\bar f$ of the oriented mixed host plus guest crystals at the
interface summed over both orientations, and $P$, the total
concentration of nuclei or seed crystals killed by the kinetic
inhibition effect and their associated chiral polarizations or
excesses. Recall that \textit{both} the hydrophobic and kinetic
effects are operative for hydrophobic amino acid additives.

As we see in Table\ref{rachydrophobic}, for all cases the interface
is covered with crystals of both orientations. This is not
surprising, since we start from a racemic composition, but no
initial ``Adam" crystal, this must lead to an equal ratio $f:{\bar
f} = 1:1$ of the enantiotopic faces exposed to the solution, thus
$od(\%) = 0$. By the same token, the $ee$ of the solution and the
net $ee_{oc}$ of occluded monomers must also be zero. There is no
symmetry breaking in this idealized situation. In the actual
experiments, floating glycine crystals can exhibit two distinct
morphologies, either pyramidal or plates \cite{Lahav1984,Lahav1988}.
The former are the result of the hydrophobic orienting effect, while
the latter morphology results from the kinetic inhibition effect.
Because of this, we can infer the crystal morphologies implicit in
our simulations: indeed, the concentrations $f$ and $\bar f$
correspond to the pyramidal form (resulting from the hydrophobic
effect) whereas the concentration $P$ corresponds to the plate-like
form (resulting from the kinetic effect).

Thus for values of $w/w_{gly}$ as low as $10^{-4}$, we found
enantiomorphous pyramids showing thus no macroscopic evidence for
the kinetic retardation of growth by for example, leucine. By
increasing the additive concentration, a \textit{morphological
change} of the floating glycine crystals is numerically observed,
from pyramids $f + {\bar f}$ to plates $P$ (inhibited seeds). The
concentration of the pyramids decreases while that of the plates
increases as we increase the concentration of the racemic additive.
This simulation result is in qualitative accord with the experiment
where an evident morphological change from pyramids to plates was
observed upon increasing the leucine concentration (i.e., the
hydrophobic additive employed there). \cite{Lahav1988}. The
initially racemic amino acid solution remains racemic at all later
times, no matter how large the initial amount of additives is.

\begin{table}[h]
\small
  \caption{\ Orientation of floating glycine crystals in the presence of mixtures of (L,D) hydrophobic additives}
  \label{rachydrophobic}
  \begin{tabular*}{0.5\textwidth}{@{\extracolsep{\fill}}llllll}
    \hline
    (L,D)\\Hydrophobic & $f+\bar{f}$ & $P$ & $od$ & $ee$ & $ee_{oc}$\\
    $w/w_{gly}$ & $mol\cdot l^{-1}$ & $mol\cdot l^{-1}$ & $(\%)$ & $(\%)$ & $(\%)$\\
    \hline
    $10^{-4}$ & $8.15 \cdot 10^{-5}$ & $4.38 \cdot 10^{-6}$ & 0 & 0 & 0 \\
    $10^{-3}$ & $6.33 \cdot 10^{-4}$ & $3.54 \cdot 10^{-4}$ & 0 & 0 & 0 \\
    $10^{-2}$ & $2.38 \cdot 10^{-3}$ & $1.51 \cdot 10^{-2}$ & 0 & 0 & 0 \\
    $10^{-1}$ & $3.48 \cdot 10^{-3}$ & $2.12 \cdot 10^{-1}$ & 0 & 0 & 0 \\
    \hline
  \end{tabular*}
\end{table}

\subsubsection{Racemic mixture hydrophobic additives + an initial oriented glycine
crystal}.

The reaction scheme in Figure \ref{SchemeII} starts from a
\textit{first random crystallization}, so we include this feature in
the initial conditions. We consider the previous situation
($ee_{0}=0$) and suppose that there is one single crystal of glycine
exposing say the (010) face toward the solution at an initial time;
this means starting with a small initial concentration of
$(f_{2,0})_0$.  Since only one initial crystal can not be taken, it
can however be "modeled" taking by a fraction of the critical
interface concentration $f_c$ of glycine crystals: $(f_{2,0})_0 = x
\, f_c$, where $x<<1$ is given in the first column of Table
\ref{evahydrophobic}. There we display the numerical results for
this particular situation, starting from an initial concentration of
(LD)-hydrophobic amino acids $w/w_{gly}=0.01$. The simulation
results show a preferential $(010)$ orientation of the glycine
crystals at the surface induced by the initial $(010)$ crystal. Even
if this preferential orientation is not exclusive $od < 100\%$, an
optically active solution is still generated, $ee = 100\%$.

\begin{figure}[h]
\centering
  \includegraphics[height=9cm]{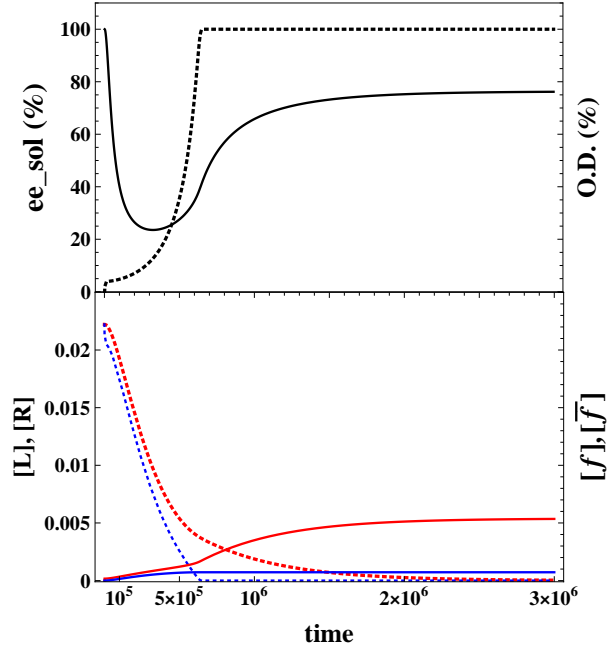}
  \caption{Racemic mixture hydrophobic additives + an initial oriented glycine
crystal. Initial conditions: $ee_0=0$, $w/w_{gly}=0.01$,
$[f_{2,0}]_0=10^{-3}\,f_c$, $ee_h0=0$, and $w_h/w_{gly}=0$. Upper
figure: dotted line represents the enantiomeric excess $ee$ of
solution and the solid line the orientation degree $od$.  Lower
figure: the dotted lines represent the concentrations of the
hydrophobic enantiomers (where [L] is the upper dotted curve), the
solid lines represent the total concentration of oriented crystals
(where [f] is the upper solid curve).}
  \label{fgr:example1}
\end{figure}

The single crystal occludes the $D$ enantiomers enantioselectively,
thus enriching the aqueous solution with the $L$ monomers. If this
small excess can induce preferential orientation of the further
growing crystals of glycine again with the (010) face pointing
toward the solution, replication will ensue by cascade mechanism
finally leading to a separation of enantiomeric
territories\cite{Lahav1988}. From Table \ref{evahydrophobic} we see
that increasing the initial concentrations of the first oriented
Adam crystals leads to increased degree of orientation $od$, and to
increased enantiomeric excesses of both the amino acids in solution
$ee$ and of those that have been occluded by the pyramidal crystals,
$ee_{oc}$. Dynamical aspects of the symmetry breaking can be
appreciated from the time evolution of the concentrations and chiral
excesses as displayed in Figure \ref{fgr:example1}.

\begin{table}[h]
\small
  \caption{\ Orientation of floating glycine crystals in the presence of mixtures of (L,D) hydrophobic additives and "one" initial oriented crystal}
  \label{evahydrophobic}
  \begin{tabular*}{0.5\textwidth}{@{\extracolsep{\fill}}llllll}
    \hline
    $(f_{2,0})_0/f_c$  & $f+\bar{f}$ & $P$ & $od$ & $ee$ & $ee_{oc}$\\
     $x$ & $mol\cdot l^{-1}$ & $mol\cdot l^{-1}$ & $(\%)$ & $(\%)$ & $(\%)$\\
    \hline
    $0.01$ & $3.02 \cdot 10^{-3}$ & $1.51 \cdot 10^{-2}$ & 30.04 & 100 & 7.10 \\
    $0.05$ & $4.59 \cdot 10^{-3}$ & $1.47 \cdot 10^{-2}$ & 62.46 & 100 & 21.55\\
    $0.1$ & $6.11 \cdot 10^{-3}$ & $1.42 \cdot 10^{-2}$ & 76.32 & 100 & 33.60 \\
       \hline
  \end{tabular*}
\end{table}

\subsubsection{Chiral hydrophobic additives}.

Instead of racemic hydrophobic additives we next consider adding a
chiral hydrophobic compound (i.e., L-$\alpha$-amino acids) to the
supersaturated solution of glycine. This means we now start with an
initial excess of the chiral hydrophobic monomers in solution so
that $ee_{0}=1$. Table \ref{chhydrophobic} shows the results for
different concentrations in terms of  the quoted $w/w_{gly}$ values.
As expected, in the presence of a resolved hydrophobic
$\alpha$-amino acid, the system achieves exclusive orientation at
the interface: $od = 100\%$ and the excess $ee$ of the remaining
amino acids in solution is $100\%$. In this situation, there can be
no amino acids occluded by the oriented crystals at the interface
since the system lacks the $D$-monomer, hence $ee_{oc}$ is not
defined. Moreover without $D$ monomers, the kinetic effect is
inoperative and so $P=0$: no plates can be formed, only pyramidal
crystals.  When L-$\alpha$-amino acids are used, these pyramids are
exclusively  $(010)$ oriented (the face exposed towards the
solution). By symmetry, D amino acids would induce the
enantiomorphous (0$\bar{1}$0) oriented pyramids\cite{Lahav1988}.
\begin{table}[h]
\small
  \caption{\ Orientation of floating glycine crystals induced by resolved L-hydrophobic additives (See Table III)\cite{Lahav1988})}
  \label{chhydrophobic}
  \begin{tabular*}{0.5\textwidth}{@{\extracolsep{\fill}}llllll}
    \hline
    (L)\\Hydrophobic & $f+\bar{f}$ & $P$ & $od$ & $ee$ & $ee_{oc}$\\
    $w/w_{gly}$ & $mol\cdot l^{-1}$ & $mol\cdot l^{-1}$ & $(\%)$ & $(\%)$ & $(\%)$\\
    \hline
    $0.01$ & $10^{-2}$ & 0 & 100 & 100 & - \\
    $0.05$ & $10^{-2}$ & 0 & 100 & 100 & - \\
    $0.1$ & $1.01 \cdot 10^{-2}$ & 0 & 100 & 100 & - \\
    $0.5$ & $1.02 \cdot 10^{-2}$ & 0 & 100 & 100 & - \\
    \hline
  \end{tabular*}
\end{table}

\subsection{Hydrophilic additives}
We now treat the general situation where both hydrophobic and
hydrophilic additives are added to the supersaturated glycine
solution. The following numerical results have been obtained for the
values $R=50$, $\gamma=0.1$, $\alpha=10^{-3}$, $\delta=10^{-6}$,
$\beta = \xi=1$, $\mu_0=10^{-6}$ and $f_c = 0.01 M$. As before, the
initial glycine monomer concentration is $[a_1]_0=4.44 \,mol \,
l^{-1}$ and for the initial glycine dimer concentration we set
$[a_2]_0=0$ and zero initial concentrations of the oriented glycine
crystals at the interface: $(f_{r,n})_0=0, (\bar f_{r,n})_0 = 0$.
Controlling the amount of hydrophilic additives gives us an
independent control of the kinetic inhibition effect.

\subsubsection{Racemic mixtures of hydrophilic additives}.

The complete amino acid solution is now composed of a mixture of
both \textit{racemic} hydrophobic $\alpha$-amino acids and by
\textit{racemic} hydrophilic $\alpha$-amino acids. Here, the role of
the hydrophobic additive is only to induce oriented crystallization
at the air/water interface. Since its composition is racemic, this
leads to an equal ratio $f:{\bar f} = 1:1$ of enantiotopic faces
exposed to the solution, and for this we use a concentration
corresponding to $w/w_{gly}=0.01$. Thus the corresponding initial
excesses of chiral hydrophobic and hydrophilic monomers are
$ee_{0}=0$ and ${ee_h}_{0}=0$, respectively. There is no initial
Adam crystal. We then observe what effect, if any, varying the
amount of the hydrophilic additives has on this initial orientation.

Numerical results showing the effect of varying the relative mass of
hydrophilic chiral monomers ($w/w_{gly}$) are shown in Table
\ref{rachydrophilic}. No matter how large the amount of hydrophilic
monomers, in the presence of racemic additives of both types, the
surface is still covered with crystals of both orientations which
immediately implies that the net excess of occluded monomers is
zero.  Note the solution achieves an extremely feeble optical
activity. In fact, these excesses are only slightly greater or at
the same level as the enantiomeric excess expected from pure
statistical fluctuations\cite{Mislow2003}. Just as reported in the
experiment, we observe how the amount of plates is much greater than
the amount of enantiomorphous pyramids: $P > (f + {\bar
 f})$\cite{Lahav1988}.

\begin{table}[h]
\small
  \caption{\ Orientation of floating glycine crystals in the presence of racemic mixtures of hydrophilic additives and 1\% hydrophobic (L,D) amino acids}
  \label{rachydrophilic}
  \begin{tabular*}{0.5\textwidth}{@{\extracolsep{\fill}}llllll}
    \hline
    (L,D)\\Hydrophilic & $f+\bar{f}$ & $P$ & $od$ & $ee$ & $ee_{oc}$\\
    $w/w_{gly}$ & $mol\cdot l^{-1}$ & $mol\cdot l^{-1}$ & $(\%)$ & $(\%)$ & $(\%)$\\
    \hline
    $0$ & $2.38 \cdot 10^{-3}$ & $1.51 \cdot 10^{-2}$ & 0 & 0 & 0 \\
    $0.01$ & $1.08 \cdot 10^{-3}$ & $2.86 \cdot 10^{-2}$ & 0 & 0 $\sim10^{-8}$ & 0 \\
    $0.05$ & $3.44 \cdot 10^{-4}$ & $3.89 \cdot 10^{-2}$ & 0 & $\sim10^{-7}$ & 0 \\
    $0.1$ & $1.86 \cdot 10^{-4}$ & $4.14 \cdot 10^{-2}$ & 0 & $\sim10^{-5}$ & 0 \\
    $0.5$ & $3.99 \cdot 10^{-5}$ & $4.38 \cdot 10^{-2}$ & 0 & $\sim10^{-7}$ & 0 \\
        \hline
  \end{tabular*}
\end{table}

\subsubsection{Chiral hydrophilic additive}.

Just as in the previous simulation, the amino acid solution is
composed of a racemic mixture of hydrophobic $\alpha$-amino acids
merely to induce the $X:Y = 1:1$ crystallization at the air/water
interface and again we use $w/w_{gly}=0.01$. But this time, we add
\textit{chiral} hydrophilic additives, and without loss of
generality we take the $L$ enantiomer. This is then a situation
described by the initial excesses of hydrophobic and hydrophilic
monomers given by $ee_{0}=0$ and $eeh_{0}=1$, respectively. Here, we
are interested in the effect of varying the relative mass of
hydrophilic chiral monomers $(w/w_{gly})$. Table \ref{chhydrophilic}
presents the numerical results for this situation. As we can see
there, the presence of a chiral hydrophilic additive induces a
preferential orientation, it is an indirect mechanism, since the
kinetic effect inhibits the growth of the crystal nuclei that would
otherwise expose the $Y$-face towards solution. Even if the
orientation is not exclusive but only preferential, the solution
does become optically active for low concentrations of the
hydrophilic additive. And from the Table we see that territorial
separation of the enantiomers is achieved.

\begin{figure}[h]
\centering
  \includegraphics[height=9cm]{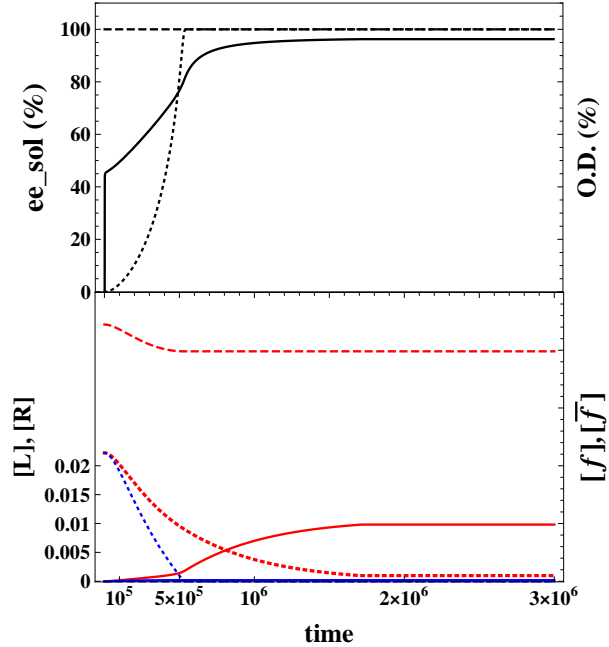}
  \caption{Chiral hydrophilic additive. Initial conditions: $ee_0=0$, $w/w_{gly}=0.01$,
$ee_h0=1$, $w_h/w_{gly}=0.01$. Upper figure: the dotted line
represents the enantiomeric excess $ee$ of the hydrophobic additives
in solution, the solid line the orientation degree $od$, and the
dashed line is the $ee_h$ of the hydrophilic additive. Lower figure:
dotted lines represent the concentrations of the hydrophobic
additives (where [L] is the upper dotted curve), the solid lines
represent the total amount of oriented crystals of each type ([f] is
the upper solid curve), and dashed line represents the hydrophilic
monomer concentration: [$\bar{L}$].}
  \label{fgr:example2}
\end{figure}

\begin{table}[h]
\small
  \caption{\ Orientation by kinetic effect of floating glycine crystals in the presence of resolved hydrophilic additives and 1\% hydrophobic (L,D)}
  \label{chhydrophilic}
  \begin{tabular*}{0.5\textwidth}{@{\extracolsep{\fill}}llllll}
    \hline
    (L)\\Hydrophilic & $f+\bar{f}$ & $P$ & $od$ & $ee$ & $ee_{oc}$\\
    $w/w_{gly}$ & $mol\cdot l^{-1}$ & $mol\cdot l^{-1}$ & $(\%)$ & $(\%)$ & $(\%)$\\
    \hline
    $0$ & $2.38 \cdot 10^{-3}$ & $1.51 \cdot 10^{-2}$ & 0 & 0 & 0 \\
    $10^{-6}$ & $2.38 \cdot 10^{-3}$ & $1.51 \cdot 10^{-2}$ & 0.51 & 100 & 0.08 \\
    $10^{-5}$ & $2.51 \cdot 10^{-3}$ & $1.51 \cdot 10^{-2}$ & 8.43 & 100 & 1.56 \\
    $10^{-4}$ & $3.28 \cdot 10^{-3}$ & $1.52 \cdot 10^{-2}$ & 38.88 & 100 & 9.21 \\
    $10^{-3}$ & $6.10 \cdot 10^{-3}$ & $1.52 \cdot 10^{-2}$ & 79.03 & 100 & 33.36 \\
    $10^{-2}$ & $1.00 \cdot 10^{-2}$ & $1.51 \cdot 10^{-2}$ & 96.26 & 100 & 76.09 \\
    $10^{-1}$ & $1.00 \cdot 10^{-2}$ & $1.51 \cdot 10^{-2}$ & 99.51 & 100 & 96.60 \\
    \hline
  \end{tabular*}
\end{table}
The hydrophilic additives have an indirect orienting effect upon the
floating glycine crystals in that they inhibit or kill the glycine
nuclei so that these are unable to occlude. Only the hydrophobic
additives can directly induce orientation. A typical time evolution
of the symmetry breaking in this situation is depicted in Figure
\ref{fgr:example2}.

\subsubsection{Hydrophobic vs. Hydrophilic additives}.

Next we study how hydrophobic and hydrophilic additives of
\textit{opposite} chiralities compete to induce a preferential
orientation of the glycine crystals at the interface. To emphasize
this competition, the amino acid solution will be composed of
strictly (L)-hydrophobic and (D)-hydrophilic $\alpha$-amino acids.
So, initially we have $ee_{0}=1$, $eeh_{0}=-1$. Take the (L)
hydrophobic additive concentration to be $w/w_{gly}=0.01$, and we
vary the (D) hydrophilic additive concentration. Table
\ref{competition} shows there is a clear ability of the resolved
hydrophobic $\alpha$-amino acids to induce a specific orientation of
the floating glycine crystals, even in the presence of large
excesses of the hydrophilic $\alpha$-amino acids of the opposite
absolute configuration\cite{Lahav1988}. This can be compared with
Table IV\cite{Lahav1988}, which established experimentally the
\textit{dominance} of the hydrophobic effect over the kinetic
effect. Note: in our model, $ee_{oc}$ is zero in this situation
since there are no $D$ hydrophobic monomers that would otherwise be
occluded.
\begin{table}[h]
\small
  \caption{\ Orientation of floating glycine crystals by hydrophobic effect in the presence of
  resolved hydrophilic additives and 1\% ($w/w_{gly}$) hydrophobic amino acids of the opposite configuration}
  \label{competition}
  \begin{tabular*}{0.5\textwidth}{@{\extracolsep{\fill}}llllll}
    \hline
    (D)\\ Hydrophilic & $f+\bar{f}$ & $P$ & $od$ & $ee$ & $ee_{oc}$\\
    $w/w_{gly}$ & $mol\cdot l^{-1}$ & $mol\cdot l^{-1}$ & $(\%)$ & $(\%)$ &
    $(\%)$\\
    \hline
    $0.01$ & $1.86 \cdot 10^{-3}$ & $4.26 \cdot 10^{-2}$ & 100 & 100 & 0 \\
    $0.05$ & $2.15 \cdot 10^{-4}$ & $4.42 \cdot 10^{-2}$ & 100 & 100 & 0 \\
    $0.1$ & $1.04\cdot 10^{-4}$ & $4.43 \cdot 10^{-2}$ & 100 & 100 & 0 \\
    $0.5$ & $2.04 \cdot 10^{-4}$ & $4.44 \cdot 10^{-2}$ & 100 & 100 & 0 \\
    \hline
  \end{tabular*}
\end{table}

\subsubsection{The amplification step}.

The hydrophobic resolved $\alpha$-amino acids are advantageous for
such a study with respect to the hydrophilic ones because both the
kinetic inhibitory effect and the stabilization of the nuclei by
hydrophobic effect act in the same direction for orientation of the
growing floating crystals of glycine. For this reason, we next study
the crystallization of glycine in the presence of partially enriched
mixtures of (L,D)-hydrophobic amino acids at various concentrations.

\begin{figure}[h]
\centering
  \includegraphics[width=0.48\textwidth]{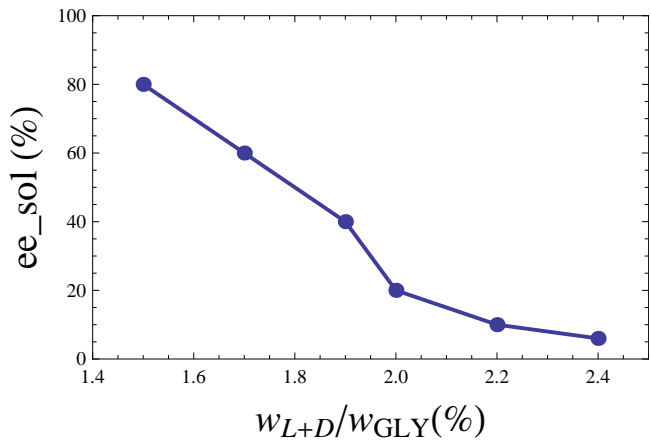}
  \caption{Correlation between the initial leucine enantiomeric excess in solution and
  the total concentration needed for the complete $(0{\bar 1}0)$ orientation of the floating glycine
  crystals. From reference\cite{Lahav1988}.}
  \label{fig3jacs88}
\end{figure}

The presence of an excess of hydrophobic (D)-amino acids will favor
the (010) oriented nucleation at the interface while at the same
time preventing the growth of ($0\bar{1}0$) oriented crystals from
the solution. The experimental\cite{Lahav1988} correlation between
the initial enantiomeric excess of the solution with the total
concentration needed to obtain complete orientation of the floating
glycine crystals is reproduced here in Fig.\ref{fig3jacs88} for
reference.

\begin{figure}[h]
\centering
  \includegraphics[width=0.48\textwidth]{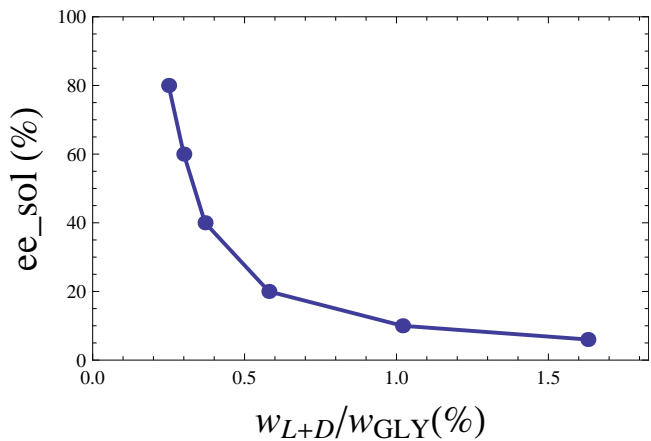}
  \caption{Correlation between the initial enantiomeric excess of the hydrophobic additive in solution and
  the total concentration needed for the \textit{maximal} $(0{\bar 1}0)$ orientation of the floating glycine
  crystals. See Table \ref{tab1tab2} and text for discussion.}
  \label{mifig3jacs88}
\end{figure}

\begin{figure}[h]
\centering
  \includegraphics[height=9cm]{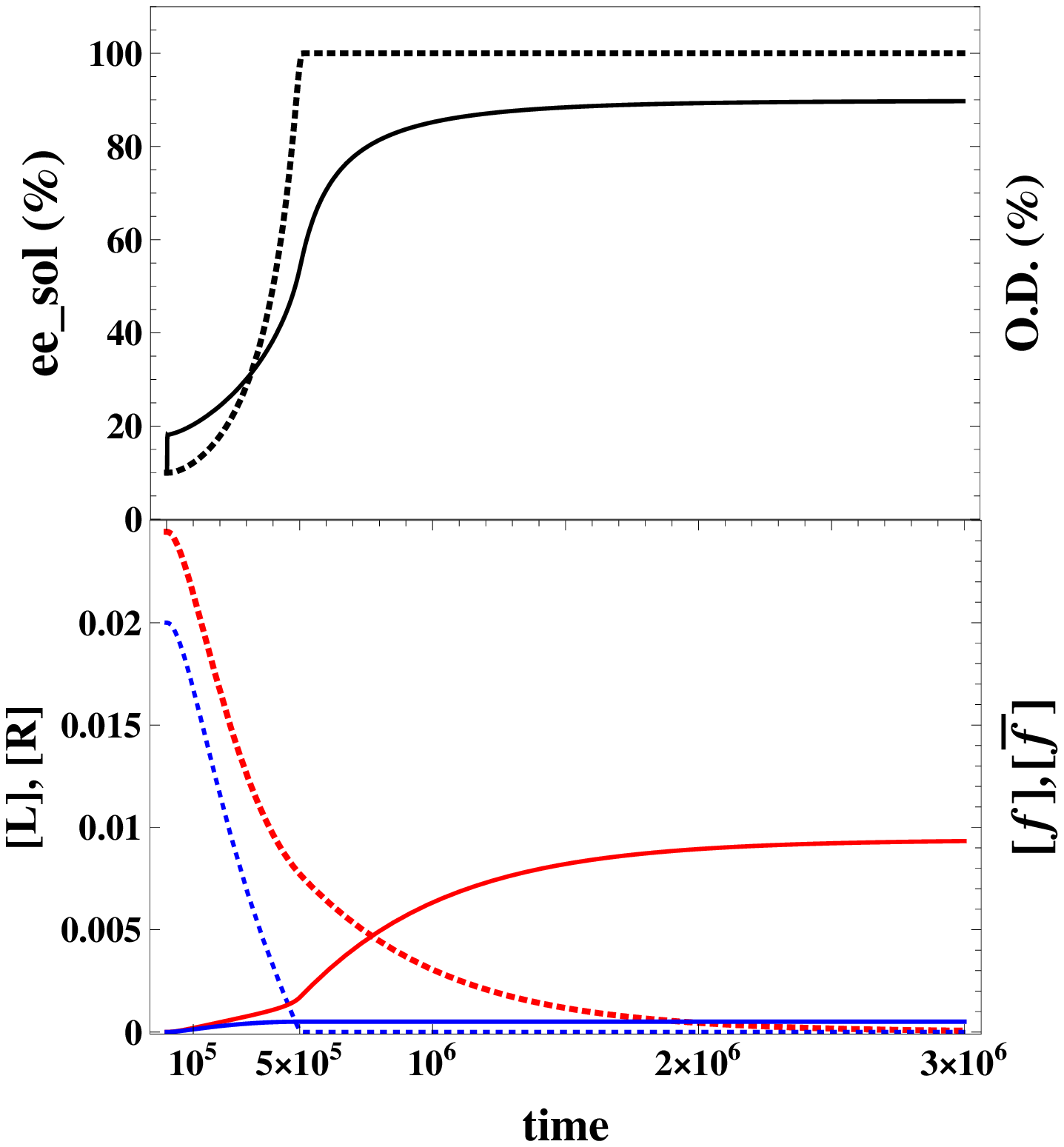}
  \caption{Partially enriched
mixtures of (L,D)-hydrophobic amino acids. Initial conditions:
$ee_0=0.1$, $w/w_{gly}=0.01$, $ee_h0=0$, $w_h/w_{gly}=0$. Upper
figure: the orientation degree $od$ (solid line) and the
enantiomeric excess $ee$ in solution (dotted line). Lower figure :
concentration of the hydrophobic monomers (the upper dotted curve is
[L]). Solid lines represent the concentration of oriented crystals
([f] is the upper curve).}
  \label{fgr:example3}
\end{figure}

The correlation between the initial enantiomeric excess of the
solution with the total concentration needed to obtain
\textit{maximum} orientation of the floating glycine crystals is
shown in Fig. \ref{mifig3jacs88}. In the Table \ref{tab1tab2} we
list all the $od$ values obtained for the different initial
enantiomeric excesses and concentrations, the numbers in boldface
correspond to the maximum orientations and these are plotted in Fig.
\ref{mifig3jacs88}. The point we wish to illustrate here is simply
that our model succeeds in capturing the general \textit{trend}
observed in the experiment, namely that smaller initial enatiomeric
excesses require greater initial hydrophobic amino acid
concentrations in order to achieve a (maximal) crystal orientation.
Figure \ref{fgr:example3} shows the temporal evolution of this
situation.

\begin{table*}
\small
  \caption{\ Orientation degree $od$ of floating glycine crystals in the presence of enriched mixture of hydrophobic additives.
  For each initial enantiomeric excess, a maximum degree of orientation is reached, to identify it we first
  locate the pair of values (the range) containing this maximum (bold font)}
  \label{tab1tab2}
  \begin{tabular*}{\textwidth}{@{\extracolsep{\fill}}llllllllll}
    \hline
    $\quad$ & $\quad$ & $\quad$ & $\quad$ & $\quad$ & $w/w_{gly}$ & $\quad$ & $\quad$ & $\quad$ & $\quad$\\
    $ee_{0}(\%)$ & 0.001 & 0.002 & 0.003 & 0.004 & 0.005 & 0.006 & 0.007 & 0.008 & 0.009\\
    \hline
    6 & 69.81 & 74.28 & 76.63 & 78.18 & 79.34 & 80.28 & 81.08 & 81.78 & 82.4\\
    10 & 79.71 & 83.18 & 85 & 86.2 & 87.09 & 87.81 & 88.41 & 88.92 & 89.38\\
    20 & 90.3 & 92.34 & 93.38 & 94.06 & 94.55 & \textbf{94.82} & \textbf{94.51} & 94.25 & 94.02\\
    40 & 96.97 & 97.77 & 98.16 & \textbf{98.28} & \textbf{98.11} & 97.97 & 97.85 & 97.74 &97.64 \\
    60 & 99.03 & 99.35 & \textbf{99.47} & \textbf{99.41} & 99.36 & 99.32 & 99.28 & 99.24 & 99.21\\
    80 & 99.77 & \textbf{99.86} & \textbf{99.88} & 99.87 & 99.87 & 99.86 & 99.85 & 99.85 & 99.84\\
    \hline
    $\quad$ & $\quad$ & $\quad$ & $\quad$ & $\quad$ & $w/w_{gly}$ & $\quad$ & $\quad$ & $\quad$ & $\quad$\\
    $ee_{0}(\%)$ & 0.01 & 0.02 & 0.03 & 0.04 & 0.05 & 0.06 & 0.07 & 0.08 & 0.09\\
    \hline
    6 & 82.96 & \textbf{85.13} & \textbf{84.26} & 83.8 & 83.52 & 83.35 & 83.24 & 83.16 & 83.11\\
    10 & \textbf{89.79} & \textbf{88.18} & 87.46 & 87.09 & 86.86 & 86.72 & 86.63 & 86.57 & 86.53\\
    20 & 93.82 & 92.7 & 92.22 & 91.97 & 91.81 & 91.72 & 91.65 & 91.61 & 91.58\\
    40 & 97.56 & 97.05 & 96.81 & 96.68 & 96.6 & 96.54 & 96.5 & 96.48 & 96.46 \\
    60 & 99.18 & 98.99 & 98.89 & 98.82 & 98.78 & 98.75 & 98.73 & 98.71 & 98.7\\
    80 & 99.84 & 99.8 & 99.78 & 99.76 & 99.75 & 99.74 & 99.74 & 99.73 & 99.72\\
    \hline
  \end{tabular*}
\end{table*}
\begin{table*}
\small
  \caption{\ Maximum orientation degree $od$ of floating glycine crystals in the presence of enriched mixture of hydrophobic additives
  for each initial enantiomeric excess }
  \label{tab1tab3}
  \begin{tabular*}{\textwidth}{@{\extracolsep{\fill}}lllllll}
    \hline
    $ee_{0}(\%)$ & 6 & 10 & 20 & 40 & 60 & 80\\
    \hline
    $w/w _{gly}$ & 0.0163 & 0.0102 & 0.0058 & 0.0037 & 0.003 & 0.0025\\
    \hline
    \end{tabular*}
\end{table*}

Experimental proof that hydrophobic effect plays a dominant role in
the orientation of the glycine crystals can be deduced from  Table
IV of reference\cite{Lahav1988}. Nevertheless, both effects are
always present for the hydrophobic amino acids: those in solution
contribute to the inhibitory kinetic effect, this is depicted in
Figure \ref{fig3jacs88}. This also explains why, at higher
concentrations, the glycine plates observed in the experiment are so
thin. At lower initial $ee$, the higher concentrations needed appear
to forbid formation of floating plates altogether\cite{Lahav1988}.

\section{Conclusions and discussion}

We have presented a mathematical model for achieving resolution of
racemic solutions of alpha-amino acids and glycine into enantiomeric
territories based on a chemical scheme proposed by the Rehovot group
some years ago. Their crystallization experiments provide a simple
model for the generation and amplification of optically active amino
acids in prebiotic conditions. The glycine and alpha amino acid
system may be relevant to the origin of optical activity as it
involves compounds that are among the simplest building blocks of
life. In this vein it is interesting to point out that a recent
prebiotic synthesis of protobiopolymers under alkaline ocean
conditions has yielded a variety of amino acids with glycine being
the most abundant\cite{Ruiz2010}. In an astrophysical context, amino
acids have been detected in room-temperature residues of
UV-irradiated interstellar ice analogues, and again glycine was
found to be the most abundant amino acid\cite{MunozCaro2002}.

Centrosymmetric glycine crystals were employed as substrates for the
total separation of occluded alpha-amino acids into enantiomeric
territories. These amino acid additives are occluded
enantioselectively through the enantiotopic $(010)$ and $(0{\bar
1}0)$ faces of the glycine crystals. Such crystals when floating at
the air/solution interface, and if properly oriented, may
incorporate only one of the two enantiomer additives present in the
solution. Complete  $(010)$ or  $(0{\bar 1}0)$ orientation is
induced by both a kinetic and a hydrophobic effect. The former
effect acts through an inhibition of the nucleation and growth of
that enantiomorph which interacts with the resolved additive. The
latter effect is due to the induction of a specific enantiotopic
face orientation (e.g., $(0{\bar 1}0)$ exposed toward the solution)
by the hydrophobic amino acids at the interface. Combination of both
effects acts in the same direction driving exclusive orientation of
the glycine crystals and thus triggers an amplification starting
from an initial random oriented crystal and a solution with low
initial enantiomeric excess. This is represented schematically by
Figure \ref{SchemeII}. Our mathematical model results from
translating this scheme into basic reaction steps and then into a
corresponding system of differential rate equations which we then
simulate to test out the hypothesized mechanisms and to underscore
the salient features of the original experiment.

Our immediate goal here is to capture the essential mechanisms
responsible for the symmetry breaking in the simplest terms
possible. This has led us to make certain simplifications, in
keeping with this aim. A major simplification is to adopt reaction
rates that are independent of the instantaneous size of the glycine
crystal and of the number of occluded amino acid guest monomers.
Another simplification is to model the processes without using
explicit coordinate dependence, which otherwise would have
necessitated the introduction of partial derivatives, diffusion
constants, spatially dependent concentrations and a coordinate
system that distinguishes the bulk three dimensional solution from
the bounding two-dimensional air/water interface or layer. There is
no doubt that these \textit{spatial} aspects play a supporting role
in the actual experiment, but they are not the primary
\textit{cause} of the mirror symmetry breaking observed there. The
fact that the coordinate-free model leads to chiral symmetry
breaking and to the territorial separation of the enantiomers is
proof of this. We have provided an analytic linear stability
analysis which gives independent confirmation of this. The
mathematical model leads to separation of enantiomeric territories,
the generation and amplification of optical activity by
enantioselective occlusion of chiral additives through chiral
surfaces of glycine crystals.

We emphasized earlier the experimentally remarkable feature that the
chemical system does away with the need for mechanical energy for
achieving the amplification of the $ee$ of the amino acids in
solution. To this end we can identify an aspect of the experiment
that serves as an effective "driving force" that maintains the
system out of equilibrium, a seemingly necessary condition for the
(permanent \footnote[6]{\textit{Temporary} breaking of mirror or
chiral symmetry is however possible even for closed systems in
thermodynamic equilibrium, see for example recent simulations of the
reversible Frank model closed to matter and energy
flow.\cite{Crusats2009,Blanco2011}}) breaking of mirror symmetry.
Indeed, the experiment requires supersaturated solutions of glycine.
To achieve supersaturation, the system must be (i) cooled down
and/or (ii) the solvent must evaporate. This then adds the element
of irreversibility.

Recent works related to the original experimental model we consider
here are summarized in a current review of the role of crystalline
architectures as templates relevant for the origins of homochirality
\cite{Lahav2011}. In brief, similar stochastic separation of
$\alpha$-amino acids has also been achieved when racemic alpha-amino
acids are occluded within enantiomorphous $\beta$-glycine crystals
\cite{Torbeev2005}. When this form is grown in either porous
materials\cite{Hamilton2008,Hamilton2009} or small solution
volumes\cite{Lee2005}, the beta polymorph glycine crystallizes as
long needles. When the $\beta$-form is grown in the presence of
DL-amino acids, the L-molecules are occluded only in one of the
$\beta$-glycine enantiomorphs, while the D's are occluded only in
the other. Thus, if a small number of one the $\beta$ enantiomorphs
crystallizes first, they will occlude only one of the alpha amino
acid enantiomers, thereby enriching the solution in the other amino
acid. This excess should prevent the nucleation of fresh $\beta$-Gly
crystals of the opposite handedness and thus lead to an enrichment
of the $ee$ of the amino acids in solution.

The present model is much simplified with respect to the actual
experimental situation in solution. It is therefore worthwhile to
identify potential future efforts aimed at ``dressing" this minimal
model so as to increase its relative ``complexity", only of course
when this is warranted or justified experimentally. One such effort
we are currently investigating is a more realistic modeling of the
self-assembly of the hydrophobic amino acid
monolayer\cite{Lahav1989} at the air/water interface and its role in
precipitating the oriented glycine crystals. Further theoretical
work aimed at incorporating other physical and chemical aspects to
the underlying scheme could also include, for example, mass
transport and diffusion, and a detailed modeling of the the glycine
crust formed at the interface. We hope to report results along these
lines elsewhere.

\section*{Acknowledgements}
The ESF COST Action CM0703 "Systems Chemistry" supported a visit of
Meir Lahav to DH at the Centro de Astrobiolog\'{i}a in Madrid and we
also thank Josep Rib\'{o} who participated in the initial
discussions and helped to foment this collaboration. We are grateful
to Meir Lahav for encouraging us to propose a mathematical model for
these experiments and for many helpful discussions and
correspondence during the course of this work. CB has a
Calvo-Rod\'{e}s predoctoral scholarship from the Instituto Nacional
de T\'{e}cnica Aeroespacial (INTA) and the research of DH is
supported in part by the grant AYA2009-13920-C02-01 from the
Ministerio de Ciencia e Innovaci\'{o}n (Spain) and forms part of the
above-mentioned COST Action. Thanks to Marta Ruiz-Bermejo for
supplying us with the references on glycine abundance in prebiotic
synthesis experiments and in simulations of interstellar ice
analogues.




\footnotesize{
\bibliography{gly} 
\bibliographystyle{plain} 

}

\end{document}